\documentclass{aa}      
\usepackage{txfonts}
\usepackage{graphicx}

\newcommand{\beq}{\begin{equation}}
\newcommand{\eneq}{\end{equation}}
\newcommand{\beqar}{\begin{eqnarray}}
\newcommand{\eneqar}{\end{eqnarray}}
\newcommand{\barn}{\begin{eqnarray*}}
\newcommand{\earn}{\end{eqnarray*}}

\def\simgt{\lower.5ex\hbox{$\; \buildrel > \over \sim \;$}}
\def\simlt{\lower.5ex\hbox{$\; \buildrel < \over \sim \;$}}

\begin{document}

\title{Non\,-Thermal Emission from AGN Coronae}

\author{G. Torricelli-Ciamponi\inst{1} \and P. Pietrini\inst{2} \and A. Orr\inst{3}}

\institute{
Osservatorio Astrofisico di Arcetri, Largo E. Fermi 5, I-50125 Firenze, Italy 
\and
Dipartimento di Astronomia e Scienza dello Spazio, Largo E. Fermi 5, 
I-50125 Firenze, Italy
\and
Research and Scientific Support Department of ESA, ESTEC, Postbus 299,NL-2200,
 AG Noordwijk, The Netherlands}

\offprints{P. Pietrini, 
\email{paola@arcetri.astro.it}}
\date{Accepted  }

\titlerunning{AGN coronae}
\authorrunning{G. Torricelli, P. Pietrini \and A. Orr}

\abstract{ Accretion disk coronae are believed to account for X-ray emission
in Active Galactic Nuclei (AGNs). In this paper
the observed emission is assumed to be due to a population of relativistic, non-thermal electrons
(e.g. produced in a flare) injected at the top of an accretion disk magnetic loop.
While electrons stream along magnetic field lines their energy distribution evolves in time essentially 
because of inverse Compton and synchrotron losses. The corresponding time dependent emission
due, in the X-ray energy range,  to the inverse Compton mechanism, has been computed. Since the typical
decay time of a flare is shorter than the integration time for data acquisition in the X-ray domain,
the resulting spectrum is derived as the  temporal mean of the real, time-dependent, emission,
as originated by a series of consecutive and identical flares. The  model outcome  is compared 
to both the broad band BeppoSAX X-ray data of the bright Seyfert~1 NGC~5548, and to a few general 
X-ray spectral properties of Seyfert~1s as a class. The good agreement between
 model and observations suggests that
 the presently proposed non-thermal, non-stationary model could be a plausible explanation
 of AGN X-ray emission, as an alternative to thermal coronae models.

\keywords{Radiation mechanisms : non-thermal -- X-rays : galaxies --
 Galaxies : nuclei -- Galaxies : Seyfert -- Quasars : general}}

\maketitle

\section{Introduction}

In these last years the idea of a hot corona lying above the
AGN  accretion disk  has been developed  in order to account for substantial
emission in the X-ray energy range. 
Indeed, the original suggestion of this scenario dates back to 1977 (Liang \& Price
\cite{liang77}), referring to the X-ray binaries' context, and to 1979 (Liang \cite{liang79}), 
more specifically regarding AGN X-ray emission.
The  hot corona can, in fact,  Comptonize
the soft photons emitted by the ``cool'' accretion disk, thus producing
high energy radiation. Several scenarios have been developed for
AGN coronae, both analytically and numerically (Poutanen \& Svensson \cite{pout}; Dove et al. \cite{dove}; 
Di Matteo \cite{dima98}; Merloni \& Fabian \cite{merlo}; Miller \& Stone \cite{misto00}).
Haardt \& Maraschi (\cite{hm91}, \cite{hm93}) analyzed in detail  and developed
the idea of a radiative coupling
between the two phases of the system (disk+corona); subsequently,
 in order to better match the observations, Haardt
et al. (\cite{hmg94})
 have developed a more detailed model of a non-uniform, blob-like  corona.
From the physical point of view, in the
above papers the hot corona is assumed     as  the a  region
where at least part of the  accretion gravitational energy is released.
The existence of an optically thin coronal structure, sandwiching the disk  and characterized by 
a  strong  magnetic coupling 
with the optically thick disk, was firstly
suggested by Galeev et al. (\cite{ga79}). The idea is that the
 accretion disk  differential  rotation, together with  the
natural presence of a magnetic field, could reproduce a situation
analogous to that present  in the atmospheres  of  convective stars.
As Heyvaerts \& Priest  (\cite{hp89}) have shown, 
loops and arcades can form also in  AGN accretion disks;
these structures, connecting remote points of the disk itself,
can convert disk kinetic energy into magnetic energy and 
subsequently  dissipate it by emitting the observed spectrum. 
Magnetic reconnection is invoked as the physical process responsible for this
last conversion, even though all the details of the process itself are not yet understood.

An issue that  has not yet been clarified is whether 
   the analogy between stellar and AGN coronae
  is limited to the above described features or whether it is
more comprehensive.
Once a physical description is chosen in which the magnetic field is  
 responsible for transferring accretion  energy from the
disk to the AGN corona, we can ask ourselves whether the magnetic field 
can act in a different way and produce a different
environment  with respect to the stellar case. 
It seems reasonable that
if the global picture of a  stellar corona is applicable in the AGN context,
the same should be true for the relevant physical processes as well.
The  analysis of solar flares  has shown  (see  Masuda et al.  (\cite{masu94})
 for a short summary)
 that  accelerated particles are responsible
for the hard ($>10$~keV) X-ray emission via the bremsstrahlung interaction
with thermal matter. Moreover, even in the absence of conspicuous
flares, X-ray brightenings and type III radio bursts show
the presence of non-thermal electron beams
in the solar corona (Klein et al. \cite{kl97}).
Hence, the existence of impulsive injection of 
non-thermal particles
seems to be a common ingredient in the physics of stellar coronae.
 The question we are interested in is whether, in the specific AGN context,
  an {\it impulsive, non-thermal} electron distribution can be responsible 
  for the emission of the 
observed X-ray radiation or, on the contrary, a stationary and  thermal electron  population
at high temperature
(like the one  adopted in the model proposed by Haardt \& Maraschi (\cite{hm91},\cite{hm93}),
and Haardt
et al. (\cite{hmg94})) Comptonizing the soft disk photons is the only possible 
explanation of the observed high energy emission.  

Stationary non-thermal models have been developed  starting from the 1980s
in order to explain AGN X-ray emission (for a review see Svensson \cite{sv94} and 
Ghisellini \cite{ghise94});
however, these models, at least in their ``standard'' versions (Svensson \cite{sv94}; Svensson
\cite{sv96}), showed some problems as for the reproduction of high energy observations of Seyfert 
spectra by OSSE,
in particular referring to the indications of high energy cutoffs in the spectra, and to 
the lack of detections at
$\sim$~0.75-3 MeV by  COMPTEL (Svensson \cite{sv96}; Haardt \cite{haardt97}). 
 In our framework, {\it impulsive} injection of the relativistic electron population is determinant
for what regards the specific characteristics of the non-thermal model we are proposing, because it 
is exactly this feature that makes the model intrinsically non-stationary. This peculiar property of our 
scenario distinguishes it from previous ``standard'' non-thermal models and will be explained 
and discussed in the following Sections.

Therefore, starting from the knowledge that we have of solar/stellar coronae
mechanisms and trying to extend it to the AGN coronae scenario, we shall investigate 
the {\it time evolution}
(see Sect.~2) and the emission   (see Sect.~3) of an  
ensemble of accelerated  non-thermal electrons  injected
in a magnetic loop. How these accelerated particles are
produced  is not relevant  to the following analysis and, on the other 
hand, it is still matter of debate even in the solar corona, although
it seems to be plausibly related to magnetic activity processes (such as reconnection).\\
In Sect.~4 we describe and discuss the application of the model that we have devised to the 
 Seyfert~1 galaxy NGC 5548; in Sect.~5 the results  of this analysis are 
compared to BeppoSAX observations. Sect.~6 is devoted to 
the discussion of the descriptive capability of the model regarding 
 to general trends and properties
identified for Seyfert~1 X-ray spectra, in order to test our model's relevance with 
respect to this type of source.
Finally, Section~7 includes both an analysis of our model as compared with previous models in
literature, and the conclusions we draw from 
the present work.

\medskip

\section{Behavior of accelerated electrons}

When a certain amount of accelerated electrons
is injected around the top of a magnetic loop, part of them
is trapped within the curved magnetic field lines,
while part of them precipitates in the denser
plasma, where the loop bases are anchored.
 The electrons which are magnetically reflected inside
the loop loose their energy
by colliding with the thermal plasma present in the loop and
by emitting  synchrotron, inverse Compton (IC) and bremsstrahlung (BR)
 radiation.
How long it takes to thermalize these electrons and how
the electron energy distribution changes in time is analyzed hereafter.
The group of precipitating electrons 
interacts with the denser plasma (of the disk or of the
photosphere) and 
stops after a rapid bremsstrahlung emission. This emission
is what we identify with the hard X-ray ($>10$~keV) emission
in solar flares.

The relative importance of the trapped and of the precipitated electrons
depends on the specific physical configuration and 
on the density gradient of thermal matter in  the loop.

An estimate of the typical loop extension in the AGN accretion disk corona context can be
obtained for instance following Di Matteo (\cite{dima98}): extending the results of 
solar atmosphere magnetic buoyancy simulations (Shibata et al. \cite{shibata89}), this author 
gives an estimate of the loop top height ($H_{\rm{flare}}$), which reads
$ H_{\rm{flare}}\sim 8 H_{\rm {disk}}$, where $H_{\rm {disk}}$ is the pressure scale height 
of the accretion disk. Furthermore, if, still following Di Matteo (\cite{dima98}), 
we remind that, for a standard Shakura \& Sunyaev (\cite{ss73}) disk, it is 
$H_{\rm {disk}}/R \la 0.1$ (where $R$ is the radial distance from the central black hole on the disk
midplane), we can relate $H_{\rm{flare}}$ to the Schwarzschild radius, 
$R_{\rm S}\equiv 2GM_{\rm BH}/c^2$; in fact, for the accreting disk central regions we can  assume
a representative  radial distance from the black hole $R\sim 10 R_{\rm S}$, and this leads
to the following approximate relation
$H_{\rm{flare}} \la  8 R_{\rm S}$  and, finally, to the estimate 
$H_{\rm{flare}} \la 2.36\times 10^{13} (M_{\rm BH}/10^7 M_{\odot})$~cm.
Other approximate estimates of coronal region extension can be found 
in the work of Miller \& Stone (\cite{misto00}), whose MHD numerical simulations of a coronal
structure formation above a weakly magnetized accretion disk indicate that, in the specific 
conditions of the simulation at least, most of the buoyantly rising magnetic energy is dissipated 
between 3 and 5 disk scale heights (i.e., what we have called $H_{\rm {disk}}$) above the disk
itself, and in the work of Liu,
Mineshige \& Shibata (\cite{liu02}), who cite a representative size value $\sim 10 R_{\rm S}$~cm.
All the above cited estimates are, in conclusion,  in substantial agreement with an evaluation 
of $H_{\rm{flare}}$ ranging from $\sim 3 R_{\rm S}$ to $\sim 10 R_{\rm S}$. 

This allows us 
to finally estimate the typical electron dynamical time in the loop, that can be defined as
$t_{\rm dyn} = H_{\rm{flare}}/c$ and represents the time it takes for the electrons to reach
the disk; we also define a characteristic radiation time, $t_{\rm emis}$, as the typical time it
takes for the
radiation emitted around 10 keV (the energy range we are mainly
interested in) to decrease by a factor of twenty in luminosity. 
  When $ t_{\rm emis} < t_{\rm dyn} $
 the electron distribution evolves on short time scales 
with respect to the dynamical timescale 
and it is depleted before reaching the
the impact region; in  this limit  the energy lost as impulsive
bremsstrahlung emission 
can be disregarded. This appears to be the case in  the AGN coronal environment,
as it will be verified in Sect.~4. 

\subsection{Electron energy loss rates} 

Given  an electron energy distribution 

\beq
n(\gamma_{\rm o},t=0)=n_{\rm o} (\gamma_{\rm o}-1)^{-s}~~~{\rm el./cm^3/}d\gamma,
\label{eldistr}
\eneq
where $\gamma_{\rm o}\equiv \gamma(t=0)$,
in the energy range (in units of $m_{\rm e}c^2$, where $m_{\rm e}$ is the electron rest mass)
$$1\leq \gamma_{\rm o1} \leq \gamma_{\rm o} \leq \gamma_{\rm o2},$$
this distribution  evolves  subject to the system energy losses. The relevant energy
loss rates for the present problem are  in principle the following
(see, e.g., Ginzburg \cite{gin}; Lang \cite{lang}; Blumenthal \& Gould \cite{blum} ):

$\bullet$ collision losses  in the relativistic limit :
$$
{d \gamma \over dt}=-1.5\times 10^{-14} n_{\rm th}[73.4+\ln(\gamma/n_{\rm th})] ~{\rm s}^{-1},
$$
where $n_{\rm th}$ is the thermal density inside the loop and it is expressed in cm$^{-3}$;
this expression can be further approximated as follows
$$
{d \gamma \over dt}\simeq -1.5\times 10^{-14} n_{\rm th}[74.4-\ln(n_{\rm th})] ~{\rm s}^{-1}
$$
\beq
~~~~~~~~~~~~~~~~~~~~~~~\equiv  A_0(n_{
\rm th})~~~ {\rm s}^{-1},
\label{collos}
\eneq
since $\ln(\gamma)$ is a small, slowly varying term
in the limit where collision losses are important (see later, eq.~(\ref{gamlim}));

$\bullet$ bremsstrahlung losses:
$$
{d \gamma \over dt}=-1.4\times10^{-16} n_{th}[ln(\gamma) +0.36] \gamma~~{\rm s}^{-1} 
$$
\beq
~~~~~~~~~~~~~~~~~~~~~~~\equiv  A_1(n_{
\rm th},\gamma) \gamma~~ {\rm s}^{-1}
\label{brlos}
\eneq

$\bullet$  inverse Compton and synchrotron losses:
\beq
{d \gamma \over dt}=-3.2\times 10^{-8} (U+ {B^2 \over 8\pi}) \gamma^2 \equiv   A_2(U,
 B) \gamma^2 ~~{\rm s}^{-1}
\label{comlos}
\eneq
where 
$B$ is the magnetic field (in gauss)
 and $U$ is the energy density (erg/cm$^3$) of soft photon radiation field present in  the
loop structure. 
 If ~$L_{\rm soft}(\epsilon$)~(erg~s$^{-1}$keV$^{-1}$) is the  seed
 (IR-UV) photon  source spectral luminosity
 coming from the AGN disk  and
 $L_{\rm local}(\epsilon$)~(erg~s$^{-1}$keV$^{-1}$) is the radiation field 
 luminosity (see eq.~(\ref{localsynch}) in Sect.~3 for its explicit definition) 
  emitted 
by the electrons of the relativistic distribution 
 through synchrotron mechanism in the local magnetic field,
then $U$~(erg~cm$^{-3}$)
can be derived    as
$$
U={1 \over 4 \pi c } F_{\rm geom}(R_{\rm UV-X})
\int^{\epsilon_{\rm Max}}_{\epsilon_{\rm min}}L_{\rm soft}(\epsilon)~d\epsilon~~+
$$
\beq
~~~~~~~~~~~~~~~~~~~~~~~~~~~~~~~{3 \over 4 \pi c R^2_{\rm em}}\int^{\epsilon_
{\rm Max}}_{\epsilon_{\rm min}}
L_{\rm local}(\epsilon)~d\epsilon~~~~~
\label{u}
\eneq
where $\epsilon$ is the photon energy, that we express  in keV,
in the range $\epsilon_{\rm min} \leq \epsilon \leq \epsilon_{\rm Max}$. 
In this expression we take $\epsilon_{\rm min}= 10^{-5}$~keV and 
$\epsilon_{\rm Max}= 0.1$~keV, $R_{\rm UV-X}$ is the mean distance between the source of 
soft photons and the X-ray emitting  loop structure and $ R_{\rm em}$ 
is the characteristic length scale of the emitting region. 
The quantity $F_{\rm geom}(R_{\rm UV-X})$
is sort of a geometric factor,  accounting for the fact that soft radiation from the disk
reaches the coronal region from different different positions in the
disk and therefore is differently ``diluted''. We adopt a simplified treatment of this effect,
following Ghisellini et al. (\cite{ghise04}) (see their eq.~(14)), so that it can be represented 
by the quantity $F_{\rm geom}(R_{\rm UV-X})$, which depends basically only on a length scale 
representative of  the vertical distance of the X-ray emitting structure from the disk 
itself, namely $R_{\rm UV-X}$.

All the above energy losses make the electron distribution change
its energy dependence. However, depending on the physical conditions of the plasma,  
their relative importance can be rather different, and in the following we discuss
this issue referring to the coronal environment of our model scenario.

First,
bremsstrahlung losses
do not seem to play an important role in the electron
distribution evolution for the typical range of $n_{\rm th}$ that we can estimate for our 
problem: they are negligible with respect to collision losses
for $\gamma<200$ and with respect to Inverse Compton and synchrotron losses 
for $\gamma > 200$, if $B>50$~gauss and $n_{\rm th}< 10^{11}$~cm$^{-3}$.
In the context of AGN coronae these limitations on the values of magnetic field and 
thermal density seem to be reasonably  met  and they do not pose very stringent 
conditions. In particular, this is true in the framework of the model we are defining for 
coronal emission, whose origin we imagine to be non-thermal. In fact, 
we have to  require that a thermal component in the coronal loops has negligible effects
on the resulting X-ray spectrum. In other words, we want to disregard  any possible thermal 
Comptonization or recoil effects on the spectrum itself. This implies  a tighter upper limit
for the density ($n_{\rm th}$) of the thermal component of the coronal loops 
(see Sect.~4.3 for a more detailed discussion and estimate for the case of NGC~5548).
In the light of the above considerations, we can therefore safely conclude that, in the context of 
AGN coronae and in particular in our model scenario, the range of plausible physical conditions 
is such that
bremsstrahlung losses are basically ineffective with respect to those due to IC and synchrotron processes.

Secondly, for what regards collision losses, we can note that 
at low energies these will make the
distribution flatter, while inverse Compton  and synchrotron losses will steepen it
at high energies. A critical value for the electron energy can be easily derived 
comparing eq.s~(\ref{collos}) and (\ref{comlos}), when $A_0\geq A_2$:
\beq
~~~~~~\gamma_{\rm lim} \equiv \sqrt {{A_0 \over A_2}} \simeq 5 \times 10^{-3} \sqrt{
{n_{\rm th} \over U+B^2/8 \pi}}.
\label{gamlim}
\eneq

 Inverse Compton  and synchrotron losses are the most efficient
 for $\gamma > \gamma_{\rm lim}$,
while for $\gamma < \gamma_{\rm lim}$ collision losses are
the most important ones. 
Adopting a representative value for the coronal structure scale length $\sim 5R_{\rm
S}$ (see Sect.~2), a rough estimate of the soft radiation energy density as 
$U\sim 0.01 L_{Edd}/4\pi c (5R_ {\rm S})^2$ gives $\gamma_{\rm lim}\sim 4$  
for $M_{\rm BH} = 10^7 M_{\odot}$ and for a 
reasonable (see Section 4.3) upper limit $n_{\rm th}\sim 10^{11}$~cm$^{-3}$ for
the thermal density. 
Furthermore, from eq.~(\ref{gamlim}) it is clear that, when $A_0\leq A_2$, IC and 
synchrotron losses will be 
dominant in any case, i.e. for any meaningful  value of $\gamma$. With the same order of magnitude 
evaluation of $U$ that we have used above, we can define an estimate of the upper limit of the
thermal density to maintain the condition $A_0\leq A_2$ fulfilled, so that collision losses 
are effectively ``negligible'' with respect to IC and synchrotron losses in the electron
distribution evolution for any possible physical value of $\gamma$. It turns out that, 
for this to occur, $n_{\rm th} < 6.4\times 10^{10}$~cm$^{-3}$.
Since for our model consistency (see below in the present Section and in Section~4.3 for more
details) we have to require a much lower thermal density in the coronal loops, 
the discussion above clearly shows  that in our
context collisional losses can be disregarded
as well.

\medskip

\subsection{The evolution of the electron distribution: an analytic solution}

Melrose \& Brown (\cite{melro76}) derived a formal method for the solution of the general 
equation for the evolution of a distribution function $n(\gamma,t)$ for particles in energy phase space.
In the limit of negligibility of 
precipitating electrons (which applies to our problem as discussed in Sect.~2)
this formal solution gets simplified. 
Taking also into account the further simplification due to the fact that bremsstrahlung   and
collisional losses 
are negligible, the remaining relevant loss rates can be expressed as proportional to 
a power of the electron energy
$\gamma$ with a coefficient $A_2$ that does not depend on $\gamma$ itself,
 but may depend
on time through the locally produced (synchrotron) luminosity, $L_{\rm local}$.
This condition 
proves quite useful, since it
makes it possible to derive, following Melrose \& Brown (\cite{melro76}) formal procedure, an  analytical
solution for the time evolution of 
the initial electron energy distribution $n(\gamma_{\rm o},0)$.

The resulting time dependent
electron distribution after a time $t$ can be expressed as :
\beq
n(\gamma,t)~~~=~~~~
 n_{\rm o} {1 \over
[1+\gamma \theta]^2}
 \left[{\gamma \over
 1+\gamma \theta} -1 \right ]^{-s}
\label{distr1}
\eneq
with
$$\theta= \int^{t} _{0} A_2(t') dt'.$$
The range of $\gamma$ over which the distribution
extends is also evolving in time and it
becomes

$$\gamma_1(t)
\leq\gamma(t) \leq 
\gamma_2(t)$$

where
$$\gamma_1(t)={\gamma_{\rm o1} \over 1-\gamma_{\rm o1} \theta }
$$
$$\gamma_2(t)={\gamma_{\rm o2} \over 1-\gamma_{\rm o2}  \theta }.$$

{\section {  The overall  emission}

From the above expressions for  the time   dependent electron distribution,
inverse Compton, synchrotron  and bremsstrahlung emission  can be computed. 
Obviously the spectral luminosity due to those radiation
processes evolves in time as well,
owing to the time evolution of the electron distribution.
As a matter of fact, in our context, the IC component is 
the sole significant contribution to the resulting X-ray emission in our model.
Indeed, relativistic bremsstrahlung emission
is not explicitly described in this Section, since  it 
turns out to be negligible in the present scenario, as we briefly discuss in the following.

In fact, we have already shown, at the end of 
Sect.~2.1, that the effects of the bremsstrahlung energy loss rate on the global evolution of 
the relativistic electron energy spectrum are unimportant with respect 
to those of the other loss mechanisms at work, for whatever value of the electron energy $\gamma$,
under the physical conditions expected for an AGN coronal structure like the one we want to model.
This clearly implies that the corresponding bremsstrahlung emission has to be negligible as well
when compared to the IC component
and this is indeed 
what we have verified from our spectral luminosity computations. This 
is important to stress, since the resulting relativistic 
bremsstrahlung would peak at energies higher than those in the typical range of the first 
scattering IC component, 
so that, in principle, it could also happen that, although small, in the very high energy range
the bremsstrahlung relative contribution to the total emission could be significant.

The effects of the synchrotron  emission
of relativistic electrons in the ambient magnetic field 
have been explicitly taken into account, as far as 
the electron energy losses (and, as a consequence, the relativistic electron population 
evolution) are concerned (see  eq.~(\ref{comlos}))
and for their
contribution to the seed  photon spectral luminosity  that is to be reprocessed
through inverse Compton mechanism.
Synchrotron emission does not show up directly in the resulting spectrum  since,
for the whole of the ranges 
of physical parameters that we have selected as appropriate  for the present context
and explored by computing the resulting spectra, it turns out that\\ 
a) due to the typical range of energies of the injected relativistic electron distribution
($\gamma_{\rm Max}\sim 10^3$), 
the synchrotron component covers a range of energies which does not belong to the X-ray domain,
and, for reasonable values of the magnetic field, peaks at energies well below the
optical and, moreover, \\
b) its emission level turns out to be below that of the observed IR-opt.-UV spectrum
 as reported in literature (see next Section). 
 
However, due to the fact that synchrotron radiation is produced in the same
place,  i.e. a coronal loop where inverse Compton reprocessing occurs, the geometrical factor 
to be accounted for 
in the computation of its energy density makes its contribution non negligible
as far as  the seed photon energy density is concerned. This contribution is
taken explicitly into account by the term $L_{\rm local}$ in eq.~(\ref{u}) and in
the following eq.~(\ref{comemis}), which defines the resulting inverse Compton 
spectral luminosity as a function of time.

Defining $E$ (in keV) as the emitted photon energy and
remembering that $\epsilon$ is the incident soft photon energy, 
the inverse Compton spectral luminosity, 
in   erg~s$^{-1}$~keV$^{-1}$,  
at a given time $t$ reads

\beq
~~~~~L_{\rm IC}(E,t)
=7.5\times 10^{-15}{E~V_{\rm loops} \over 4 \pi c} \times
\label {comemis}
\eneq
$$\int^{\epsilon_{\rm  Max}}_{\epsilon_{\rm  min}}\int^{\gamma_2
}_{\gamma_1}
g(\gamma,\epsilon,t)
\left[F_{\rm geom} {L_{\rm soft}(\epsilon)
 \over \epsilon^2}+ {1 \over R^2_{\rm em}}
 {L_{\rm local}(\epsilon,t) \over \epsilon^2} \right ]
 ~ d\gamma d \epsilon ,
 $$
where
$$
g(\gamma,\epsilon,t)=\cases{  
{2~n(\gamma,t) \over 3\gamma^2}\left (1-
{E\over 4 \gamma^2 \epsilon} \right )~~~~~~~~{\rm for}~~~ 
{E\over 4 \gamma^2 \epsilon}<1\cr
0 ~~~~~~~~~~~~~~~~~~~~~~~~
{\rm for}~~~~~~ {E\over 4 \gamma^2 \epsilon}\geq 1.},
$$
In the above expression,  $V_{\rm loops}$ is the total volume of the part of the corona
where flares take place (contributing to the emission we observe) and
$L_{\rm soft}(\epsilon)$ is the illuminating IR-UV spectral luminosity 
from the accretion disk defined over the energy 
interval $[\epsilon_{\rm min}, \epsilon_{\rm Max}]$ (see Sect.~2.1). 
We assume this  soft  spectrum as known, and we model it as a parameterized 
power-law
with a high energy cut-off, as it is explicitly specified below:
\beq
L_{\rm soft}(\epsilon)= H_0 \epsilon ^{1-\rm UV2} e^{-\epsilon/E_{\rm cut}} ~~~~~~~~
\epsilon _{\rm min}< \epsilon.
\label{uvspec} 
\eneq
In this expression  UV2 is the photon index of the spectral distribution,
$E_{\rm cut}$ is  the cut-off energy (in keV), and 
$L_{\rm soft}$ is expressed in erg~s$^{-1}$~keV$^{-1}$. 
 Therefore,  for each AGN
source we intend to study, first we shall have to appropriately choose the parameters of 
our representation of $L_{\rm soft}(\epsilon)$, so as to fit the
 observed IR-UV spectrum, as described in literature.

The other term present in eq.(\ref{comemis}) , $L_{\rm local}$, takes into
account any production of soft photons by the relativistic electron distribution,
namely synchrotron emission in the loop magnetic field.
As outlined above, for our parameter ranges, 
synchrotron emission is too weak to be relevant in the observed spectrum,
but its contribution as seed radiation  for the IC reprocessing is non negligible.
The synchrotron spectral luminosity of each single active loop 
is given by:
\beq
L_{\rm local}(\epsilon,t)=2.8\times 10^{-5} {V_{\rm loops}\over q}~B ~\times~~~~~~~~~~~~~~~~~~~~~~~~~~~
\label{localsynch} 
\eneq
$$
~~~~~~~~~
\int_{\gamma_1} ^
{\gamma_2} d\gamma~n(\gamma,t) {\epsilon
\over \epsilon_{\rm
c}} \int^{\infty}_{\epsilon /  \epsilon_{\rm
c}} K_{5/3}(\xi) d \xi~~~~~~~~ {\rm erg~s}^{-1}~{\rm keV}^{-1}, 
$$
where
\beq
\epsilon_{\rm c} =1.7\times 10^{-11} \gamma^2B~~~~~~~~~{\rm keV,}
\eneq
$q$ is the number of active loops in the coronal structure (assumed constant in time)
and $ K_{5/3}$ is the modified Bessel function of order 5/3. 

Expression (\ref{comemis}) for the resulting  IC spectral luminosity takes into account 
only the first inverse Compton
scattering of the soft photons illuminating the loop, in the hypothesis that 
the second and the higher order ones 
have negligible effects on the resulting high energy spectrum in our context. 
The validity of this approximation will be discussed and verified 
in Sect.~5.

It is also important to stress that, even if this model does not belong to the
``family'' of thermal models, it shares with them the existence of a feedback between the flaring
corona and the cold underneath disk. In fact, we assume that roughly one half
of the inverse Compton  emitted radiation is directed inwards, to the disk surface; this
approximate fraction of the X-ray radiation will therefore be somehow reprocessed by the disk 
material (absorbed or ``reflected''), thus implying some dependence of the disk physical conditions on the 
coronal emission and possibly some sort of relation between the high energy coronal emission itself
and the soft disk emission.
The expected complex interplay
between disk and corona has been widely discussed by Haardt
et al. (\cite{hmg94}).
However, it is important to stress that, in accordance with Uttley et al. (\cite{uttley})
conclusions, in this work we do not demand that soft radiation from the disk is only due to reprocessing 
of the X-ray coronal emission.

{\section {Application of our model to AGN X-ray sources: the example of NGC~5548}}

The choice of adequate observational data for testing our model is based
on several criteria. Seyfert~1 galaxies (Sy 1) are excellent candidates
because a substantial amount of their observed X-ray emission is believed
to originate directly from the inner central regions of the accretion disk
(with some reprocessing in the form of reflection and absorption).
The contribution from the jet to the observed flux is believed to be
negligible in this type of AGN.

 To exemplify the applicability of our model and to fully illustrate our procedure in its
details, we have to choose a particular source and test the model itself against the source's
 observed X-ray spectrum. This is the subject of the present Section. 
However, the plausibility and validity 
of the emission model we present must be evaluated on the basis of a more general comparison 
with the observed properties of X-ray spectra of Seyfert~1 active nuclei. This comparison is
postponed to Section 6, in which the descriptive capabilities of our model will 
be  analyzed in order to account for a few general observed trends and 
properties of Seyfert~1 X-ray spectra. 

For the present purposes of quantitative and detailed comparison between theory and observations, it
is important
to choose a bright source for which broad-band X-ray
spectra are available as well as good optical/UV data. For the X-ray data
the choice of BeppoSAX is obvious, because of its unparalleled broad band
and sensitivity (Boella et al. \cite{boella97a}). Ideally, the optical/UV and X-ray
data should be simultaneous with the X-ray data, however such data sets are
extremely rare.

NGC~5548 is a well-known and nearby (z=0.017) Seyfert 1.5 galaxy, bright
in the X-ray domain.
 It was observed several times by BeppoSAX
  (see Nicastro et al. \cite{nica};  Petrucci et al. \cite{petru};
 Pounds et al. \cite{pounds}) and
a vast amount of optical and UV observations are available for it.
It is among the 8 brightest type 1 Seyfert galaxies observed by
BeppoSAX ($F_{2-10~{\rm keV}} \ge 4 \times 10^{-11}$~erg~cm$^{-2}$~s$^{-1}$).

BeppoSAX observed NGC 5548 at 3 different epochs: August 1997, December 1999
and July 2001. The source was at it brightest during the 1999 and 2001
observations. The July 2001 data set was selected for our study
because the exposure at low energies (LECS instrument) is better than in
1999. We do not have simultaneous multi-wavelength data for this source
(extending from the IR to the X-ray range) and, as a consequence, we rely on
literature knowledge of the IR-UV spectrum in what follows.

\vskip 0.5truecm

\subsection{The adopted soft photon spectrum}
The parameters
describing its IR to UV spectral luminosity  in the simplified representation that 
we have introduced for this quantity in the previous section (see eq.~(\ref{uvspec}))
have been selected by  analyzing the results described in the literature
as follows.
At energies higher than $\epsilon_{\rm min}=10^{-2}$ eV (124 $\mu$)
we assume
 UV2~$ = 1.3$ and a normalization luminosity, $L_{\rm soft}(2.25~$eV$)
= 2.8\times 10^{46}$ erg s$^{-1}$ keV$^{-1}$, derived from Ward et al. (\cite
{ward}). Since the source is highly variable this value for the luminosity
seems a good compromise between the higher value quoted by Malkan \& Sargent
(\cite{malkan}), $L_{\rm soft}(2.25$~eV$)
= 3.6\times 10^{46}$ erg s$^{-1}$ keV$^{-1}$,  and the luminosity range reported
by Wamsteker et al. (\cite{wamst}), $L_{\rm soft}(2.25$~eV$)
= 1.3 \div 2.9 \times 10^{46}$ erg s$^{-1}$ keV$^{-1}$.
The value  H$_0 = 50\;\;
{\rm km \;s}^{-1}{\rm Mpc}^{-1}$ has been assumed to derive
the above luminosities.
 Note that  between  $\epsilon_{\rm min}$ and $\epsilon_1=1.8$~eV (0.7 $\mu$)
the observed spectrum, like the one shown in Fig.~\ref{obs1}, can
be represented by a power-law with photon index  UV1~$= 2$ (Carleton et al. \cite {carle}).

An exponential cut-off at $E_{\rm cut}= 0.015$~keV, modifies the slope of the UV spectrum at high energies.
The value has been chosen so that 
the energy spectrum decreases for $\lambda < 1060 \AA$
in accordance with Brotherton  et al. (\cite{broth}).
In this way the resulting seed spectrum is also in accordance
with the EUV data reported in Chiang et al. (\cite{chiang})
where $L_{\rm EUV}(0.163$ keV)= 8.2$\times10^{43} \div 9.5\times 10^{44}$ erg~s$^{-1}$~keV$^{-1}$.

\begin{figure}
\resizebox{\hsize}{!}{\includegraphics{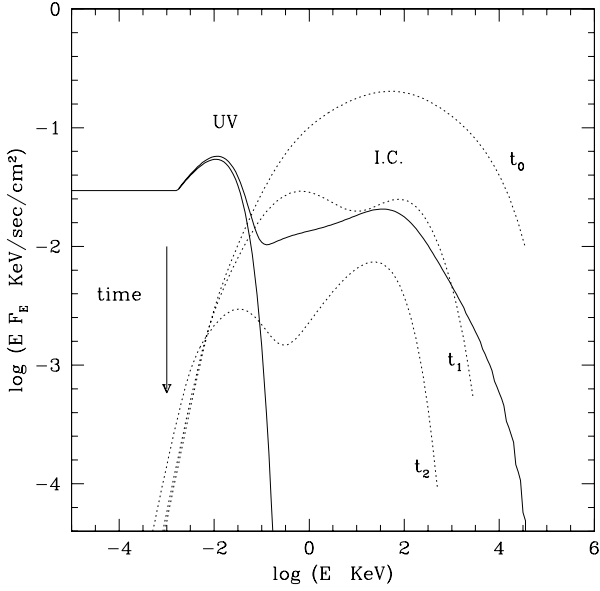}}
\caption{Dotted curves show the  inverse Compton spectrum 
obtained at  different times ($t_0=0.02$ s,
$t_1=1.0$~s, $t_2=5.0$~s 
 after the injection of an electron 
distribution characterized by $\gamma_{\rm o1}=50$, $\gamma_{\rm o2}=1000$, $s=3.0$ 
 and $n_{\rm o}V_{\rm loops}=2.57\times 10^{53}$ el/d$\gamma$ (see the text for this parameter's
 definition).  The continuous curve drawn for energies 
 $E\leq 0.1$~keV shows the adopted representation of the soft photon spectrum 
for the case of
 NGC 5548. 
The other continuous curve,
 extending up to very high energy, is the spectral energy distribution of the time averaged
resulting emission, as derived from eq.~(\ref{mean}).}  
\label{obs1}
\end{figure}

\subsection{The definition of our X-ray model spectrum: an illustration}
In the following, unless otherwise specified, we translate 
the results of our calculations of spectral luminosities  into 
the corresponding spectral energy distribution $Ef(E)$~(keV~cm$^{-2}$~s$^{-1}$), 
which is commonly used and directly comparable to observations, 
through the obvious relation 
$Ef(E) \equiv EL(E)/(4\pi D^2)$, where $D$ is the distance to the chosen 
source, appropriately evaluated. 

Fig.~\ref{obs1} represents an illustrative summary of the results of our model, as it will be extensively
explained in the following.
In fact, this figure shows different curves referring to the outcome of the spectral 
luminosity computations of our model (as defined by eq.~(\ref{comemis}))
for the case of NGC~5548 that we have chosen as a test of the model itself. 

The specific meaning of the various curves in Fig.~\ref{obs1} is explained hereafter.
The soft photon (IR-Opt.-UV) spectral energy distribution resulting 
from the observational data in the literature outlined above for NGC~5548 
is shown in Fig.~\ref{obs1} as the continuous curve 
extending up to $E=0.1$~keV.
The dotted curves represent the contribution to our model spectrum due to inverse 
Compton emission at three different times 
$t_i = 0.02$, 1.0, 5.0~s  
(after the high energy electron injection in the loop), 
computed from expression (\ref {comemis}). 
It is apparent that the 
 spectral energy distribution strongly depends on time 
and, in the X-ray energy range, it decreases with time, due to the  evolution of 
the relativistic electron population, i.e. to its depletion starting from the high energy 
end of the distribution.

For the results shown in Fig.~\ref{obs1}, the initial physical parameters of 
the relativistic electron distribution 
injected in the loop at $t=0.0$~s 
are those specified in the caption of Fig.~\ref{obs1}.  Here we introduce as a significant parameter
the product $n_{\rm o}V_{\rm loops}$, with $n_{\rm o}$ the normalization constant 
in relation (\ref{eldistr})
defining the electron energy distribution, and $V_{\rm loops}$ the total volume occupied by the
active loops; its physical meaning is such that $(n_{\rm o}V_{\rm loops}d\gamma)$ represents the number
of accelerated electrons in the whole flaring volume in an energy interval $d\gamma$. 
In addition, the other two physical parameters entering 
the emission computations (see eq.~(\ref{comemis})) have been chosen as 
 $R_{\rm UV-X}= 1\times 10^{14}$~cm,  $n_{\rm th}= 
3\times 10^{8}$~cm$^{-3}$; we note in passing that these values are 
 those for which the best fit of the spectrum of NGC~5548
 is obtained, as described in the following.

There is one more curve appearing in Fig.~\ref{obs1}, i.e. the continuous one extending
over the whole X-ray energy range, and in the following we discuss its meaning 
and the way it is computed within the framework of our model.
From Fig.~\ref{obs1} it is apparent  that the computed emission changes on short time scales.
Especially around a few hundred keV
 the decrease of the   spectral energy distribution value 
with time is significant, owing to the depletion of high energy
electrons due to inverse Compton and synchrotron losses.
Note that we have defined  the quantity $t_{\rm emis}$ as the time  the spectrum
at $E=10$ keV takes to decrease by a factor of  twenty with respect to 
its initial value.
For the case shown in Fig.~\ref{obs1} 
  a value of $t_{\rm emis}$ around  4.0~s can be derived.
It is evident that every observational procedure
which takes a time $> t_{\rm emiss}$ to get a spectrum 
will basically make a temporal mean of 
the time dependent spectra in the data acquisition.
If X-ray emission in AGN is due to such a  mechanism
the fast evolving flare spectra imply that a rapid succession of flares
 must take place over different parts of the disk so that when one flare is fading out
another one  starts brightening. If the decay time for a flare is
$t_{\rm emis}$, we need one  flare to  start
at least every $t_{\rm emis}$ seconds.  We assume here a uniform
distribution in time of flares in order  to reproduce the observed emission.
This would not be the case for a real AGN where the observed
variability can be due to a non uniform distribution in time 
and in intensity of  the flares. However, this scenario does not reduce the
applicability of the model, since the observational procedure does introduce
an averaging operation on the emitted spectra on longer time scales. 
In conclusion, what we observe is a temporal mean
of the emitted spectra, i.e., in terms of spectral luminosity,
\beq
L_{\rm MEAN}(E)= {1 \over t_{\rm emis}} \int^{t_{\rm emis}}_0 L_{\rm IC}(E, t) dt.
\label{mean}
\eneq
The result of this temporal mean in terms of an energy spectrum ($Ef(E)_{\rm MEAN}$)
is drawn in Fig.~\ref{obs1} as the continuous curve extending up to high energies.

The fast time evolution of the IC emission has important
consequences. 
In fact, since the  estimated loop length scale is (see Sect.~2) $H_{\rm flare} > 3R_{\rm S}\sim 
6.2\times 10^{13}$~cm for $M_{\rm BH} = 7\times10^7 M_{\odot}$, 
 the following relationship holds  
$$t_{\rm dyn} \simeq H_{\rm flare}/c > ~ 2.07\times 10^3~{\rm s}> t_{\rm emis}.$$
The numerical evaluation above refers to the specific case of NGC~5548, but, when repeating 
the same estimate for a  lower value of the central black hole mass, say $10^7 M_{\odot}$, 
we get the same ordering of the relevant timescales, namely
$t_{\rm dyn} \simeq H_{\rm flare}/c > ~ 300~{\rm s}> t_{\rm emis}$.
Therefore, we can generally conclude that we do not need  to take into account
that at each electron reflection inside the magnetic loop 
a  certain fraction  of the electron beam
precipitates into the disk, as explained at the beginning of Sect.2.

\subsection{Parameters of the model and their role}
 The mean spectrum shown in Fig.~\ref{obs1} depends on many free parameters:
the scale length of the corona, $R_{\rm UV-X}$, the electron distribution 
parameters, $\gamma_{\rm o1}$, $\gamma_{\rm o2}$, $n_{\rm o}V_{\rm loops}$ and $s$, 
the parameters entering 
the electron losses, namely the magnetic field $B$ and the plasma thermal density $n_{\rm th}$.
As for the other scale length appearing in the equations relevant to the definition of the model
spectrum, that is $R_{\rm em}$, the characteristic linear size of each single emitting loop region,
in our scenario it is directly related to the global corona scale length parameter defined above, namely
$R_{\rm UV-X}$, by means of our specific choice of simplified geometrical structure; in fact, 
supposing that approximately a fraction $w = 1/10$ of the total coronal volume is flaring, i.e. 
is responsible for X-ray emission, at any given time, and that the  number of simultaneously 
active loops (emitting flares) is 
$q$, the relationship between the global coronal region dimension and   the scale length of each
emitting structure (flare) $R_{\rm em}$ is the following
\beq
 R_{\rm em}\simeq R_{\rm UV-X}~{\left (w\over q\right)}^{1/3}.
\label{looplength}
\eneq
This choice is, of course, somewhat arbitrary, but nevertheless in its simplicity sufficiently 
general; in the fitting procedure described and discussed in the following Sections we have chosen
for the number $q$ of simultaneous flares at any given time a representative value of 10, as
inferred by Haardt
et al. (\cite{hmg94}), but this parameter can be changed
and the related consequences are briefly discussed in Section~6. 

The IR-UV spectrum which influences both the electron losses (see eq.~(\ref{comlos}) and
(\ref{u})) and the IC emission (eq.~(\ref{comemis})) is given as known  for the time being,
even though its time variability makes the reliability of this assumption not 
obvious when simultaneous
UV and X-ray observations are not available.
We have computed the mean spectrum, like the one shown in Fig.~\ref{obs1}, changing in turn
one of the above listed parameters. The effects each of them has on the
mean spectrum are the following:

\medskip
a) changing $R_{\rm UV-X}$  changes the shape of the mean spectrum in a complex 
way, since this parameter appears not only explicitly in eq.~(\ref{comemis}), but also, implicitly, 
through the evolution of the electron energy distribution, which depends on IC losses (see
eqs.~(\ref{comlos}),(\ref{u})); hence, 
the value of $R_{\rm UV-X}$ 
is determined by the comparison with the observations 
within a suitably selected range;

b) a decrease in $\gamma_{\rm o1}$ makes the spectrum peak (which is around
$100$ keV in Fig.~\ref{obs1}, computed for $\gamma_{\rm o1}= 50$) shift towards lower energies;
 
c) changing $\gamma_{\rm o2}$ does not have any significant effect on the 
temporal mean of the spectrum 
for $E \leq 300$~keV,  since the high energy portion of the 
electron distribution is in any case depleted very fast;

d) $n_{\rm o}V_{\rm loops}$ is essentially  the normalization parameter of the resulting 
spectrum for $E > 0.1$ keV (in other words, it is somewhat proportional to the energy spectrum 
level) and  the comparison with observations determines its value;

e)  an increase in the value of $s$ makes the spectrum steeper for
$E>100$ keV;

f) $B$  has a complex influence on the mean spectrum; however
 equally good fits of the data can be recovered for different
values of the magnetic field intensity; 

g) $n_{\rm th}$ does not change the shape of the mean spectrum for
$E \leq 10^5$ keV.

In conclusion, the parameters which determine the spectrum in the range of BeppoSAX
observations (0.1-200 keV) are $R_{\rm UV-X}$,
 $n_{\rm o}V_{\rm loops}$, $\gamma_{\rm o1}$ and $s$. In our study these
parameter values will be defined by the fit to the observational data 
of BeppoSAX for NGC~5548. 
On the other hand, the values of the other three parameters  $\gamma_{\rm o2}$, $B$ 
and $n_{\rm th}$ are chosen by us before the fitting procedure. 
We set $\gamma_{\rm o2}= 1000$, since we have 
verified that this has no influence on the physics
of the model. 

As for the other two input parameters, $B$ and $n_{\rm th}$, the following 
general considerations help us  to estimate reasonable values.
Assuming a value  $\simeq 7\times10^7~ M_{\odot}$ for the central black hole mass of NGC~5548
(Wandel et al. \cite{wande}), the resulting Schwarzschild radius  is 
$R_{\rm S} \simeq 2.07\times 10^{13}$~cm and an estimate of the lower limit for the length scale 
of coronal structures can be derived as 
$R_{\rm UV-X}>3~R_S\simeq (R_{\rm UV-X})_{\rm min}$ (see Sect.~2). 
 This lower limit for the length scale of the corona implies in turn an upper
limit for the thermal density of the background thermal coronal plasma. In fact, 
as already mentioned in Sect.~2.1, 
it is our requirement that the thermal plasma in the coronal structure does not 
affect significantly the resulting high energy spectrum; this can be translated into a condition
on the optical depth to scattering of the thermal component, $\tau_{\rm T}$, that must be
$\tau_{\rm T}=n_{\rm th} \sigma _{\rm T} R_{\rm UV-X} << 1$. 
To meet this condition a limitation on the thermal density is obtained: 
$n_{\rm th}<<{1 \over \sigma_{\rm T} R_{\rm UV-X}} \la {1 \over \sigma_{\rm T}
 (R_{\rm UV-X})_{\rm min}} \simeq (\sigma_{\rm T} 6.2\times 10^{13}~{\rm cm})^{-1}
 \simeq 2.4\times 10^{10}$~cm$^{-3}$. We just note here that for lower $M_{\rm BH}$ 
 we expect a scaling of the limit on $n_{\rm th}$, but for $M_{\rm BH}\ga 10^7 M_{\odot}$ it
 would anyway be $n_{\rm th} \ll 10^{11}$~cm$^{-3}$.
 In our model we choose $n_{\rm th}=3\times 10^8$~cm$^{-3}$
which  substantially fulfills the condition above, $\tau_{\rm T} \ll 1$,   
on a linear length scale $R_{\rm UV-X}<10^{15}$ cm and, as we shall see in the next Section, 
``a posteriori'', certainly  is in 
accordance with that same condition for 
the value of $R_{\rm UV-X}\simeq 1\times 10^{14}$~cm that we find from 
the fitting procedure on  NGC~5548.
It is important to mention here that no specific evaluation of the temperature 
of this thermal component has been done, nor it is the aim of this work to do it. We only
can say that, since we assume that most of the energy discharged in the coronal loop goes into
accelerating  the relativistic electron population, we do not expect (actually we can exclude)
that this temperature can be as high as the values estimated in the thermal Comptonization models
for the X-ray coronal emission. On the contrary, it is plausible that 
the thermal plasma in the loops thermalizes  at comparatively low temperatures, with respect to those
typical of thermal Comptonization  models for X-ray emission. This inference, even though 
purely qualitative, together with the requirement expressed above of small optical  depth
to electron scattering of the thermal plasma,  
is the reason why we can neglect  any thermal component 
effects [i.e., no thermal comptonization
effects, no recoil distortion of the high energy end of the X-ray spectrum (see Krolik \cite{krolik99})].

As far as the magnetic field is concerned, its value is limited by energetic considerations.
In fact, in the working hypothesis that magnetic energy conversion is the source
of the accelerated electron energy, the following relationship must be fulfilled
$${B^2 \over 8~ \pi} > n_{\rm o} \int^{\gamma_{\rm o2}}_{\gamma_{\rm o1}} \gamma_{\rm o}
 (\gamma_{\rm o}-1)^{-s} d\gamma_{\rm o}. $$
In this expression the electron distribution parameters are derived from the
fit, as described above, and hence an iterative procedure is required to find
a suitable value for the magnetic field, since $B$ itself enters the fit.

For NGC~5548 we have found that  values of $B$ in the range $\sim 110-500$~gauss 
can produce model spectra that do fit the observed X-ray spectrum, provided that the typical coronal
scale length $R_{\rm {UV-X}}$ is suitably chosen: it turns out that smaller coronal structures
must be associated with stronger magnetic fields, in order to produce similarly fitting model
spectra.

\bigskip

{\section {Comparison with observations of NGC~5548 }}

 The very extended energy range of BeppoSAX data makes it
extremely well suited for comparisons with our model.

We have thus computed the mean luminosity spectrum, $L_{MEAN}(E)$,
for the case of NGC~5548 using the IR-UV parameters listed in the previous
section. Then we have generated a grid of our spectra using the {\tt XSPEC}
{\sc atable} format in order to compare our model to observations.

NGC 5548 was observed by BeppoSAX from 2001 July 08, 08:35 UT, to July 11,
  00:16 UT. We used BeppoSAX data obtained with the Low-Energy
Concentrator Spectrometer (LECS; 0.1--10~keV; Parmar et~al. \cite{parmar}), 
the Medium-Energy Concentrator Spectrometer (MECS; 1.65--10~keV;
Boella et~al. \cite{boella97b})  and the Phoswich Detection System (PDS; 15--300~keV;
Frontera et~al. \cite{frontera}). The net exposures in the LECS, MECS and PDS
instruments  are 42~ks, 98~ks and 48~ks, respectively. The net count
rates for the LECS, MECS and PDS instruments are  0.37 cts s$^{-1}$,  0.55 cts
s$^{-1}$ and  0.90 cts s$^{-1}$ respectively.

The BeppoSAX data were reduced using the SAXDAS 2.3.0 data analysis package.
Good data were selected from intervals when the elevation angle above the
Earth's limb was $>4^{\circ}$ and when the instrument configurations were
nominal. The standard PDS collimator dwell time of 96~s for each on- and
off-source position was used together with a rocking angle of 210 arcminutes.
LECS and MECS data were extracted using radii of 8 arcminutes and 4 arcminutes,
respectively. Background subtraction for the LECS and MECS were performed
using the standard background files. The PDS background was estimated
from the offset field according to the standard procedure.

The LECS and MECS spectra were rebinned to oversample the full width half
maximum (FWHM) of the energy resolution by a factor 3 and to have,
additionally, a minimum of 20 counts per bin to allow use of the
$\chi^2$ statistic.
The PDS data were rebinned using the standard logarithmic binning recommended
for this instrument. Data were selected in the energy ranges
0.13--2.4~keV (LECS), 1.65--10~keV (MECS), and 15--200~keV (PDS),
where the instrument responses are well determined and sufficient
counts obtained.
\vskip 0.5cm

\subsection {Fitting procedure}
First of all we have fitted these data with our model + Fe K$\alpha$ line+ Compton 
reflection hump in the energy range $3 \div 200$~keV, by allowing  the
following free parameters to vary: 
$R_{\rm UV-X}$ and $n_{\rm o}V_{\rm loops}$.
The fitting process has been repeated for different values of the parameters 
$\gamma_{\rm o1}$,  $s$, in order to determine for which values the fit on 
$R_{\rm UV-X}$ and $n_{\rm o}V_{\rm loops}$  best reproduces the spectrum (see Section~5.2).
The reflection hump component is modeled  by appropriately using {\sc pexrav} 
in {\tt XSPEC} (where
the required input power-law with cutoff parameters are derived by recursively ``fitting'' 
this simple representation of the X-ray primary spectrum to our IC  model spectrum).
The narrow Fe K line is simply modeled with a {\sc zgauss} additive component.

We start the fitting procedure over this limited energy range, excluding the ``low'' 
energy portion of the data, because we want at first to disregard the possible effects of 
a warm absorber, 
which, at low energies, 
could easily hide our model's spectral features.
Over this energy range we obtain a very good fit ($\chi_{\nu}^2 =
 0.78$ (d.o.f. =  88)),
once we have chosen the values of the parameters of the injected electron 
distribution at time t=0, i.e.  $\gamma_{\rm o1}=50$, $\gamma_{\rm o2}= 
1000$, $s=3$ (see Sect.~5.2).
Fig.~\ref{obs2} shows a comparison of our best fit model to the data in this ``medium-to-high''
energy range.
For the physical parameters 
directly defining the generation of our non-thermal IC model spectrum, we obtain the values
listed in Table~1; the results of the fit for those parameters referring to the other components 
of the observed X-ray spectrum, such as
the Iron line and the reflection hump, are not included in Table~1, and are described 
later.
The parameter $N_{\rm tot}$ appearing in Table~1 is defined as 
$$N_{\rm tot}= n_{\rm o}V_{\rm loops} \int^{\gamma_{\rm o2}}_{\gamma_{\rm o1}}
(\gamma-1)  ^{-s} d\gamma~;$$
this parameter represents the total number of relativistic electrons in the coronal structure 
volume and it
is a function of $\{n_{\rm o}V_{\rm loops}, \gamma_{\rm o1}, \gamma_{\rm o2},s\}$.
We want to stress again that the $\chi ^2$ value  in this
case is very good, 
confirming that our model gives a good interpretation of the data in this energy range, 
which basically corresponds to the energy interval over which the intrinsic IC model 
strongly influences and determines the overall resulting spectral shape, 
together with the reflection component, which, in turn, depends 
on the intrinsic spectrum itself. 

As already noted above, Table~1 only shows the parameters of our IC model spectrum, 
but of course the fit  determines 
also the parameters referring to the other components of the resulting X-ray spectrum that we
included, that is the Fe K line and the reflection hump.
Our best fit selects a central energy for the line 
$= 6.37\pm 0.08$~keV and a line width $\sigma < 170$~eV. 
As for the reflection hump, 
the resulting reflection factor
$R\equiv\Omega/2\pi$, where $\Omega$ is the solid angle subtended by the reflecting matter
(see Magdziarz \& Zdziarski \cite{magdz}), is 0.8, and the cosine of the inclination angle
of the reflecting slab-like material with the line of sight
has been fixed at 0.9, like in Pounds et al. (\cite{pounds}), and Perola et al. (\cite{perola02}).
It is worth noting that, if the cosine of the inclination angle is left free to vary, 
a value  $\chi^2_{\nu} =  0.80$ (87 d.o.f.) is obtained, 
corresponding to a value 0.8 for the cosine itself and again R = 0.8, 
with the parameters 
reported in Table~1 essentially unchanged.

\begin{table}[ht]
\caption[]{The 3-200 keV physical parameters for our model, selected as described in the text.}

\begin{tabular}{lll}
\hline\noalign{\smallskip}
 $\gamma_{\rm o1}$ & $=$ &  50 \\
 $\gamma_{\rm o2}$ & $=$ &  1000 \\
 $s$ & $=$ & 3.0 \\
 $B$ & $=$ & 500~gauss\\
 $R_{\rm UV-X}$ & =  & $(1.0_{-0.01}^{+0.05}) \times 10^{14}$ cm \\
 $N_{\rm tot}$ & =  & $(5.34_{-0.02}^{+0.03}) \times 10^{{49}}$ electrons \\
  $\chi^2_{\nu}$ & $ = $ &   0.78 (88 d.o.f.)\\
\noalign{\smallskip\hrule\smallskip}
\end{tabular}
\end{table}

\begin{figure}
\resizebox{\hsize}{!}{\includegraphics[angle=-90]{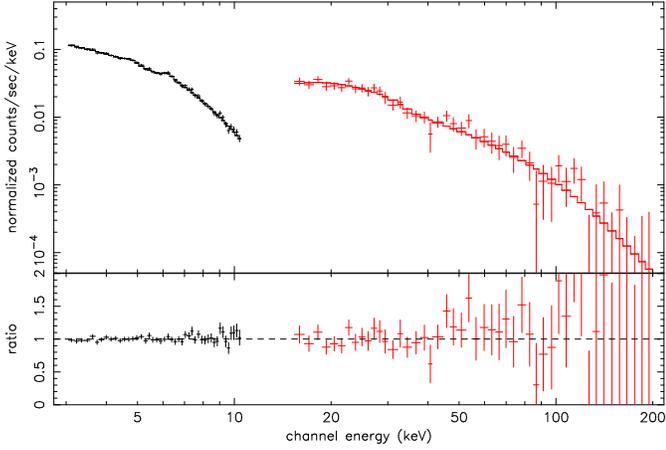}}
\caption{BeppoSAX spectrum of NGC 5548 and the best fit with our model + Fe~K$\alpha$ +
 reflection hump, in the range 3-200 keV (see text).} 
\label{obs2}
\end{figure}

We then use these
same parameter values, by fixing them in the {\tt XSPEC} procedure,   to compare 
the model (that we have found well fitting the 3-200~keV range) to the whole 
energy range of BeppoSAX data; the result of  this comparison is  much worse, 
giving an unacceptably large value of $\chi_{\nu}^2$ ($\sim 16$).
In fact, keeping fixed  the parameters selected in the restricted 
energy range (3-200~keV)  and ``freezing'' the normalization constant of the LECS data group to
a value such that the ratio of LECS/MECS normalization lies within the acceptable interval (see
Fiore et al. \cite{fiore}) and, at the same time, such as to make the model match the data in 
the low energy range ($E \la 0.4$), the model flux $f(E)$ (keV~cm$^{-2}$~s$^{-1}$~keV$^{-1}$)
that we obtain
is instead significantly higher than the unfolded observed spectrum 
in the range 0.4- 3 keV; this is shown in Fig.~\ref{obs2b}.

\begin{figure}
\resizebox{\hsize}{!}{\includegraphics[angle=-90] {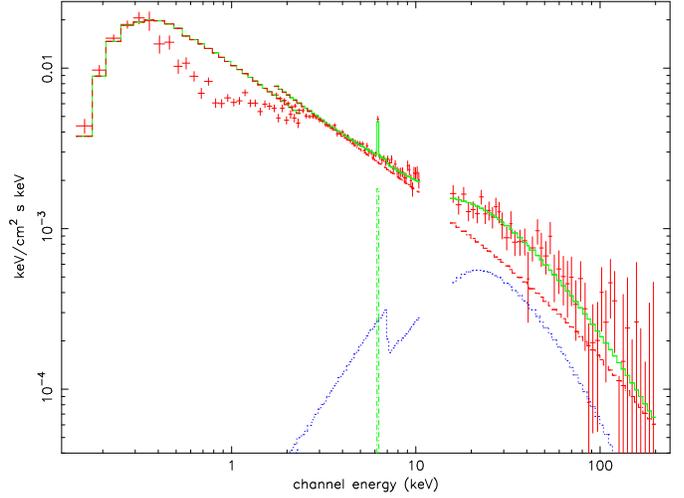}}
\caption{Comparison of BeppoSAX data over the whole energy range observed  with the 
IC ``primary'' spectrum + Fe line + reflection component as derived from the 3-200~keV
best fit: the necessity of some absorption component is apparent from the plot; see the text
for further explanation.}
\label{obs2b}
\end{figure}

This  energy range is exactly the one where the expected effects of a warm absorber are
most evident (Nicastro et al. \cite{nica2}).
Indeed, recent observations of this energy range with the unprecedented 
spectral resolution  of XMM-Newton, have shown an astounding complexity of spectral features, 
emission and absorption lines, possibly blueshifted, (see Steenbrugge et al. \cite{steen};
Kaastra et al. \cite{kaastra}) 
revealing how complicated and composite a realistic description of the absorbing gas 
must be, certainly falling beyond the scope of the present work. In the context of 
Beppo-SAX data analysis in the ``low-energy''
spectral region, characterized by a limited spectral resolution (with 
respect to that of the grating spectrometers
on board of XMM), similarly to what is done by Pounds et al. (\cite{pounds}), 
we choose to give an approximate description of the warm 
absorber effects with a series of absorption edges, namely 5 edges. 

We postpone the discussion of these 
edges until Section~5.3 and at present we only say that including them  
in the fitting procedure 
(each of them represented by a  multiplicative component {\sc zedge} in {\tt XSPEC}),
and keeping the IC model spectrum parameter  values fixed as those in Table~1, now 
we obtain a much better fit ($\chi ^2_{\nu} = 0.96$ (133 d.o.f.))
over the entire observed energy range 0.12-200 keV; 
this is  shown in Fig. \ref{obs3}.

\begin{figure}
\resizebox{\hsize}{!}{\includegraphics[angle=-90] {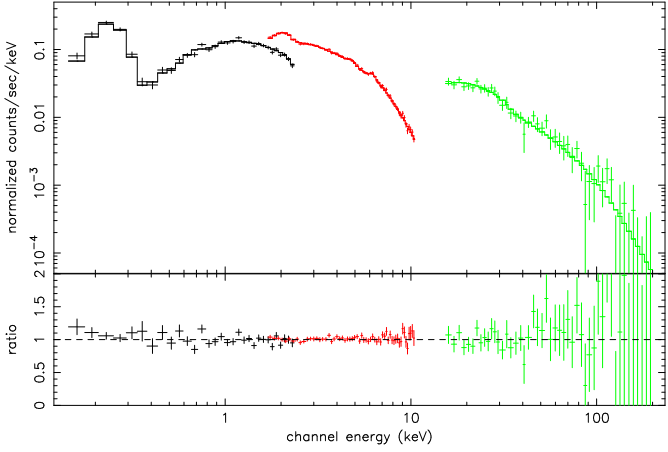}}
\caption{BeppoSAX spectrum of NGC 5548 and the best fit with our model + Fe~K$\alpha$ +
 reflection hump +  absorption edges (see text). No soft excess is included.} 
\label{obs3}
\end{figure}

\subsection {Physical parameters for the generation of the IC spectrum}

Table~1 shows the best fit values for the parameters defining the 
physical scenario that we have devised for the generation of the primary X-ray 
spectrum. 

We have also tried to fit models corresponding to different
values of $s$ and $\gamma_{\rm o1}$; the ones we have finally chosen, given in Table~1, 
are those for which the best fit is obtained, i.e. corresponding 
to one of the lowest obtainable $\chi^2$ values. 
However, it is interesting to note that variations of these two 
parameters within $\sim15\%$ of the reported values  result in a fit which is still a reasonable one, 
if not slightly better: 
for example, for $s=2.5$ the final $\chi^2$ value turns out to be even lower than the one found 
for $s=3$ (see Table~1). Nevertheless, we have preferred and shown the fit obtained for the model 
corresponding to $s=3$, since in this
case the decrease of the spectrum at high 
energies is steeper than in the case corresponding to $s=2.5$.
Regarding the energy range of the injected electron distribution, recent work
on electron acceleration due to magnetic reconnection related processes (Lesch \& Birk \cite{lesch}; 
Schopper et al. \cite{schop}) in the AGN context and, even more specifically, in the AGN corona context,
has shown that such processes guarantee acceleration up to maximum energies around 
$\gamma \sim 2000$, which is essentially in accordance with our choice of the upper limit of the energy 
range of the initial injected electron distribution.

The values of $R_{\rm UV-X}$ and  $N_{\rm tot}$
 reported in Table 1 have been directly derived from the best fit procedure. 
In the hypothesis that approximately one tenth of the coronal volume 
is flaring at the same time and that the emission 
is due to a series  of $q=10$ almost simultaneous flares, the characteristic linear scale of a typical
emitting region can be estimated as 
$R_{\rm em}\simeq R_{\rm UV-X}/(10q)^{1/3}  \simeq~ 2.15\times10^{13}$~cm 
and the corresponding  minimum variability time scale is $\sim~ 7.2\times10^{2}$~s, in 
accordance with the the data reported by Chiang et al. (\cite{chiang}). 
In this framework the density of non-thermal electrons can be evaluated as  
 $$n_{\rm rel}\simeq {N_{\rm tot} \over 10(4\pi/3)R_{\rm em}^3}\simeq 
 1.28\times 10^{8}~{\rm cm}^{-3}$$
and the associated optical depth to (non-thermal relativistic) electron scattering is 
$\tau_{\rm rel}\equiv n_{\rm rel}\sigma_{\rm T}R_{\rm em}\simeq 1.8\times 10^{-3}$.
 With such a low optical depth, secondary inverse Compton scattering effects can be 
safely neglected.
The computation of how many double scattered photons attain $E=300$~keV
shows that their number is significantly lower than that of the photons 
that have been scattered only once.

Finally, we want mention that a similarly good fit from our model can be obtained also
for larger values of $R_{\rm UV-X}$, and, correspondingly smaller  magnetic field values, as 
noted at the end of Section~4.3. In particular, for $B= 120$~gauss, from the fit in the 3-200~keV
range, we obtain  
  $\chi^2_{\nu} = 0.76$ (87 d.o.f.), for 
 $R_{\rm UV-X} =  (6.0_{-0.01}^{+0.03}) \times 10^{14}$~cm  and 
 $N_{\rm tot}=  (1.40_{-0.01}^{+0.01}) \times 10^{51}$~electrons,
 with essentially the same resulting parameters for the iron line, and with $R=0.8$
 and  $(0.67_{ -0.13}^{ +0.13})$ for the cosine of the inclination angle for the {\sc pexrav} 
 component, defining the reflection component of the spectrum. Extending the fit to the whole 
 energy range of BeppoSAX data in the same way described in the previous section, 
we obtain  $\chi^2_{\nu} = 0.95$ (132 d.o.f.), thus giving 
as good a fit as the one described in the previous Section. For this case 
$R_{\rm em}\simeq  1.29\times 10^{14}$~cm, so that 
the relativistic electron density is
 $n_{\rm rel}\simeq  1.56\times 10^{ 7}~{\rm cm}^{-3}$, 
and the associated optical depth to scattering is 
$\tau_{\rm rel}\simeq 1.3\times 10^{-3}$.
Nevertheless, we have chosen to fully illustrate 
the fit corresponding to $R_{\rm UV-X}= 1\times 10^{14}$~cm
in the previous and in the present Sections, 
following the outcomes of the discussion of Section~2 regarding the estimate of  the typical 
scale length of active loops, which indicate  that  smaller  size coronal loops are favoured.

\subsection {Other components of the observed X-ray spectrum }

We want to mention here that the fit we have obtained seems to indicate that no significant soft 
excess is required from the 2001 data we have analyzed. This might seem in disagreement with 
Pounds et al. (\cite{pounds}) (who detect an ``apparently unambiguous excess of soft X-ray
flux below 0.7~keV''); however, this is only apparent, since their conclusion comes from the 
use of a simple power-law to describe the primary spectrum. On the contrary, in our case the X-ray primary 
spectrum itself is the result of inverse Compton reprocessing of soft photons by our evolving 
non-thermal distribution of relativistic electrons, and, as such, turns out to be characterized 
by a more complex behaviour. Specifically, it shows a slight 
hump right in the energy region around and below $\sim$~1~keV, as it is suggested 
by Fig.~\ref{obs1} and could be better seen with a more appropriate scale. 
This is just the time-averaged effect
of the evolution of  the lower limit of the electron energy distribution, $\gamma_1=\gamma_1(t)\leq
\gamma_{\rm o1}$, and it turns out to be essentially enough 
to suppress the necessity of an additional soft excess
component.
 We note, however, that we might indeed ``force'' a soft excess component in the framework of
our model and still obtain a reasonable fit; in other words, we do not {\it exclude} the
possibility of a soft excess component, but we stress again that it turns out to be not 
necessary to introduce it within our modeling scenario.

Let us briefly discuss the {\tt XSPEC} components that 
we have added to our resulting IC spectrum. In fact, we have introduced these
components since it is well known that they do have an influence on the ``low'' 
X-ray energy range for what regards the warm absorber, and in the medium-to-high energy 
range for the Iron emission line and the reflection hump (Nicastro et al. \cite{nica}; Petrucci et al.
\cite{petru}); 
as a consequence, to fit the whole BeppoSAX energy range it is necessary to account for them. 
 However, we want to stress that we  are not 
interested here in discussing the nature/origin of these reprocessing components in detail, 
but instead we want to put emphasis on the feasibility of an ``intrinsic'' emission model 
such as the one we have envisaged. Furthermore, the parameters we obtain from the general
fit for the warm absorber and for the other reprocessing components are not in disagreement 
with the values cited in pre-XMM analyses in the literature (Nicastro et al. \cite{nica}; 
Pounds et al. \cite{pounds}).

As for the five absorption edges through which we give a simplified description of the warm 
absorber effects, 
four of these edges have been fixed to the theoretical 
OVII, OVIII and NeIX/X and MgX/XI edge energy values
(and these are essentially the four edges mentioned by Pounds et al. (\cite{pounds})).
 The fifth edge, which is 
required if we want to obtain a good fit, turns out to be placed at $\la 0.5$~keV, and it could 
 correspond to the deep absorption edge of CIV predicted by photoionization models of the 
warm absorber (see Nicastro et al. \cite{nica2}, \cite{nica3}).
However, what is of interest  to our present purpose is that we do obtain a good fit 
by allowing for absorption edge   energies  and resulting optical depth values 
which are in accordance  with published studies of 
warm absorbers  prior to the high energy resolution observations of XMM. In fact, 
this allows us to devote our attention to the ``primary'' X-ray continuum
 as emitted in the non-thermal scenario we have devised.
Moreover, it is not very meaningful any longer to dwell on the meaning and origin of ``low-energy'' 
spectral features as they appear from BeppoSAX data, since CHANDRA and XMM, allowing for 
much higher spectral resolution and therefore much more detail, have now provided a much more complex 
picture, revealing a richness of lines  and spectral features whose analysis requires a different approach
and different means of study, and, possibly, showing no evidence at all for the oxygen edges
previously identified in lower energy resolution observations (Kaastra et al. \cite{kaastra00}).

\section{ Application of the model to AGN X-ray sources: general considerations} 

In the previous section we have shown that the X-ray spectrum of the specific Seyfert~1 source
NGC~5548 can be interestingly well explained by the non-thermal and intrinsically non-stationary 
model that we have devised. Now we turn our attention to the analysis of the descriptive capability
of the model with respect to a few general observed trends and  properties of Seyfert~1's X-ray spectra.
We do not get into the detailed and quantitative description of a fit, like we did  
in the previous two sections; in fact, 
since in our model the reprocessed spectrum depends on the input soft radiation spectrum,
and on the loop magnetic field through the locally generated synchrotron component, which
represents a seed component for inverse Compton reprocessing as well,
to compare its outcome with the average Seyfert~1 X-ray
spectrum, we should have defined an averaged 
opt.-UV spectrum to be reprocessed  and an average magnetic field,
 and then  fit the resulting X-ray spectral distribution to the ``averaged'' observed
spectrum, obtaining values of the physical parameters coming into play that would define 
``average'' physical properties of the emitting regions. As a matter of fact, especially because of
the significant dependence of these parameters on the physical condition of the very central region
of the AGN, i.e. on the effective value of the central black hole mass ultimately, a procedure of
this kind would, in our opinion, lead to not very meaningful results. 
On the contrary, we point our attention to the capability of our model to reproduce 
a very general spectral characteristic of the observed Seyfert X-ray spectra, that is the
relatively small dispersion of the 
distribution of the spectral index $\alpha$ of a power-law like description 
of the ``primary'' spectral flux ($ f(E)\propto E^{-\alpha}$)  at least in 
the energy range $\sim 2-10$~keV 
around an average value $\alpha\sim 0.9$ (Nandra \& Pounds \cite{nandra94}; Nandra et al. 
\cite{nandra97}; Zdziarski et al. \cite{zdz95}; Zdziarski et al. \cite{zdz00}; 
Gondek et al. \cite{gondek96}; Matt \cite{matt01} for BeppoSAX broad-band observations),
by now considered as a ``canonical'' value. 
It is appropriate to note here that, strictly speaking, such an inference  on the primary 
spectrum  is somewhat model dependent, since the ``canonical'' value $\alpha = 0.9$ is 
obtained fitting the observed spectrum with a series of components (primary and reprocessed, see
also Section~5.1)
that are a priori supposed to be descriptive of the general mechanisms at work in the definition 
of the resulting X-ray radiation. In particular, the existence of a reflection component turns out 
to be determinant; for an interesting, and in a way ``historical'', analysis of the 
developments of the global phenomenological understanding of X-ray spectra of Seyfert~1
nuclei, we refer for example to Nandra \& Pounds's (\cite{nandra94}) work on GINGA data, showing 
that the strong flattening of the spectrum  in the 10-18~keV range, with respect to the  
2-10~keV behaviour, was well accounted for by introducing a reflection component in the spectral
fitting, thus giving a relatively narrow distribution of values 
of the spectral index of the power-law description of the primary spectrum 
around an average value $\sim 0.9$.\\
Back to the point, it is our purpose to show that our 
model can quite generally reproduce a spectrum that fairly well approximates a power-law behaviour 
in the range 2-10~keV with a spectral index that does match the relatively narrow range of values around 
$\alpha = 0.9$ inferred from the general interpretation of Seyfert X-ray spectra observations.  
In fact, we restrict our analysis to the energy interval delimited by $\sim$ 2 and 10~keV  respectively 
because of two orders of reasons, which we explain in the following and are anyway intimately connected
with each other. Indeed, on one hand 
our model primary spectrum, when considered on a much broader X-ray energy range (such as the one observed
by BeppoSAX), necessarily (i.e. due to its origin) is much more complex than a simple power-law.  On
the other hand, within the general interpretative framework, with the exception of the generally
narrow (see Bianchi et al. \cite{bianchi04}) Iron line at $\sim 6.4$~keV,
the effects  of the reprocessing components usually accounted for in the  theoretical explanation of 
the resulting broad-band X-ray spectrum are expected to be much more conspicuous outside of this 
$\sim 2-10$~keV range,
(because of absorption on the lower-energy side, and of reflection on the higher-energy side).
Therefore, outside of the $2-10$~keV range,
 deviations of our model spectrum from a simplistic power-law behaviour are mixed up with 
reprocessing features/distortions  and  can be disentangled only by altogether fitting  the observed
spectrum, possibly giving sort of different results for the reprocessing component parameters (with
respect to the simple primary broad-band power-law description) but in no way rejectable on a
theoretical basis.
Keeping in mind the discussion above, 
to attain our goal, we have chosen the following procedure. First, we had to define some sort of 
``average'' properties of a typical Seyfert~1 spectrum in the optical-UV range. Several works 
in the literature attempt to define these properties for the general category of radio-quiet AGNs, 
or for QSOs only (see for example Risaliti \& Elvis \cite{risaliti04} and references therein), 
but, in order to select an average opt.-UV spectral behaviour more specifically referring 
to the class of Seyfert~1s, we referred to Koratkar \& Blaes's (\cite{korat99}) work, specifically to
their Fig.~1, in which the authors show an average SED (spectral energy distribution)  
for Seyfert~1s, distinguishing it both from 
radio-quiet quasars (QSOs) and radio loud ones. Approximately fitting a opt.-UV Sy-1 spectrum from
Koratkar \& Blaes's (\cite{korat99}) Fig.~1, we computed a grid of resulting primary X-ray 
spectra from our model, changing the physical parameters that we found more significant in the 
detailed fitting of NGC~5548's X-ray BeppoSAX spectrum as performed in Section~5, namely $R_{\rm
UV-X}$ and $N_{\rm tot}$, together with the magnetic field in the active region, $B$.
We find that, for a reasonable range of $N_{\rm tot}$ values, and for a given $R_{\rm UV-X}$, 
 we can quite easily define a range of values of the magnetic field $B$ corresponding to 
 which  our model spectrum in the range 2-10~keV can be well described as power-law like, with 
 spectral index $\alpha$ values between 0.85 and 0.95. To quantify our assertion, for example, 
 this typical slope range for the power-law best describing our spectrum can be obtained 
 for $B \simeq 100-200$~gauss, when $R_{\rm UV-X}= 6\times 10^{14}$~cm, whereas 
 for $R_{\rm UV-X}= 1\times 10^{14}$~cm
 this same $\alpha$-range is reproduced for $B \simeq 400-600$~gauss.
 This is in accordance with the trend that we have briefly discussed at the 
 end of Section~4.3, referring to the specific fitting of NGC~5548's BeppoSAX spectrum; in fact,
 we do find that, even ``starting'' from an an average opt.-UV spectrum to be IC reprocessed
 by the coronal relativistic electrons, 
 an increase of the coronal structure extension requires smaller 
values of the magnetic field to 
 reproduce the canonical power-law slope, i.e. to better fit observed spectra.  

Another issue we have to confront with is X-ray variability, its general 
behaviour and its relationship with variability properties at
other  wavelengths.
X-ray spectral variability studies have shown that many
Seyfert spectra soften when the continuum flux increases
(Nandra et al. \cite{nandra97bis}; Markowitz \& Edelson \cite{mark01};
Markowitz et al. \cite{mark03}).  Also in NGC~5548
temporal variability has been observed at several wavelengths;
in particular,  UV-EUV and simultaneous X-ray variability data are available
(Chiang  et al. \cite{chiang}; Haba et al. \cite{haba}; Kaastra et al. \cite{kaas2004}),
and the same trend, namely an
UV-EUV intensity increase larger than the X-ray one, is present.
Another characteristic behaviour which appears both in NGC 5548 (Chiang  et al.
 \cite{chiang}; Kaastra et al. \cite{kaas2004}) and also in some
 other sources (see, e.g., McHardy et al. \cite{hardy}; Blustin et al.
\cite{blustin}) is the fact that
hard X-ray intensity variations show a delay with respect to the soft X-ray
ones.

A detailed comparison of our model results with the observed variability
properties is out of the purpose of this paper. However, it is important
to  understand whether our model could reproduce the most relevant 
variability  trends  outlined above. 
In our model both an increase in the magnetic field intensity and a different geometry,
i.e. less numerous but more extended active loops, can easily
reproduce a softening of the spectrum. More specifically, an increase
of the magnetic field intensity of 25\% can induce an increase in the
observed spectral luminosity of 22\% at 0.3~keV and of 7\% at 3 keV, while
a decrease of the number of active regions from ten to three  can induce an increase in the
observed spectral luminosity of 28\% and 10\%, respectively  corresponding to the same 
two energy values cited above. A non-homogeneous magnetic field showing
 different geometrical configuration and intensity in different regions of the accretion disk,
owing to the disk rotation and to its intrinsic temporal variability,
would produce changes such as the ones mentioned above in the coronal magnetic structure 
and in the number and properties of flaring loops.
This picture is in accordance with the
idea of Kaastra et al. ( \cite{kaas2004})
regarding the possible presence of a large  hot active region above the disk.
In this context,
 magnetic field properties can be imagined to evolve in time in such a way to justify the
observations of  the  hard X-ray part of the spectrum 
still increasing,  while the soft one decreases, in 
accordance with the reported time lag of hard X-ray radiation.

\medskip
\section {Discussion and Conclusions}

In this paper we have developed the idea that AGN coronae
resemble the stellar ones, where flares powered by magnetic reconnection
inject accelerated electrons into  magnetic, loop-like, structures.
In this framework,  X-ray emission from AGNs
is due to the non-thermal, time-evolving distribution of relativistic
electrons and the dominant emission mechanism is inverse Compton scattering
of IR-UV soft radiation emitted by the underlying accretion disk. The model has
been tested on one of the brightest
Seyfert~1 galaxies, NGC 5548.

Despite making exceedingly simplifying assumptions, our model gives a
very good fit to the X-ray spectrum of NGC~5548 (see Figs.~\ref{obs2} and \ref{obs3})
as observed by BeppoSAX in July 2001,
 with physically reasonable values of the model parameters.
In addition, the spectral index of the initial electron energy distribution, $s$, that we
derive turns out to be quite in accordance with the values inferred for
solar flares, hence supporting the idea that magnetic reconnection is related to the electron
acceleration mechanism.
Moreover, our  $R_{\rm UV-X}\simeq 4.8 R{\rm _S}$  is substantially consistent with
the characteristic size of
a coronal structure for an object like NGC 5548. If ten flares are active at the same time
 in the corona (see Sect.~5), the estimated size of the blob-like structure
implies that significant variations in the observed
luminosity are on time-scales longer than $\sim  7.2 \times 10^{ 2}$ s,
an estimate which is in accordance with X-ray 
luminosity variations observed up to now (Chiang et al. \cite{chiang};
Nicastro et al. \cite {nica}).
Since our model is based on a time-evolving emission
process, possible variations of the  number of active flares
and of their intensity allow to decouple X-ray intensity variations from the UV ones
and also to  reproduce the observed changes in the emission spectrum
(see section 6).
Another advantage of our model is that it does not require the introduction
of a soft X-ray emission component, since the model itself reproduces
 the so called ``soft excess'' (see Sect.~5.3).

A qualitative comparison with general trends inferred from observations of the X-ray 
spectra of Seyfert~1 active nuclei has been discussed in Section~6: both the ``average''
spectral index $\alpha\simeq 0.9$ generally determined for power-law approximations of the 
``primary'' spectrum and the variability patterns observed for several sources (and in particular
for the one, NGC~5548, we analyzed in more detail) briefly described in the previous Section 
can be rather easily reproduced in the framework of our model. This is encouraging in what concerns
the feasibility of the scenario we propose in the present work.

As outlined in the introduction,  two  main different families of models 
have been developed in order to understand the physics of the X-ray
emitting plasma: those in which the energy distribution of the electrons responsible for 
X-ray emission is thermal and
those in which it is instead non-thermal. 
It is interesting to highlight the most important  points
distinguishing the presently proposed scenario from the above cited families of models.
Differences with thermal models are straightforwardly identified, since our model is 
based on the impulsive injection
of highly relativistic electrons non-thermally distributed  in energy
(and evolving in time due to their 
energy losses), interacting with seed soft photons to produce a time dependent X-ray spectrum 
through a single-scattering inverse Compton process. On the contrary, the usual 
thermal models are characterized
by a steady-state trans-relativistic thermal energy distribution  for the electrons, 
which interact with the soft radiation field again through the inverse Compton mechanism, but 
in a multiple scattering process, thermal Comptonization, with each one of the scatterings 
involving a relatively small amount of energy exchange. 
In our scenario we do allow for an ``underlying'' thermalized coronal
 plasma component, 
 but  its physical properties differ from those of ``thermal models''
 quite significantly, as discussed in Sect.~4.3; in particular, the  density 
 of the thermalized component is much lower
 than that estimated for thermal models.
As a consequence, since the optical depth to scattering of this component is much smaller than
one,
 no significant effect from the Compton
interaction between thermal matter and photons can be expected, not even Compton ``recoil''
(see Sect.~4.3).
The  ``primary'' spectrum is therefore only determined by the non-thermal relativistic electron
interactions with the radiation field.

To this respect, it is also important to stress the properties of
 our coronal emitting component,  i.e., of the non-thermal
relativistic electron population. In fact, also  this component is optically
thin  to scattering,  since its Thomson optical depth
is very low  ($\sim 0.0018$ for the case of NGC 5548), and this justifies our choice of 
a simple single-scattering description.

Concerning the 
``standard'' stationary non-thermal models 
(see Svensson \cite{sv94}; Ghisellini \cite{ghise94}), 
two main differences with respect to our present model
can be identified {and are discussed in the following; however, it is important to 
notice that these same two differences also apply to the 
comparison with stationary
thermal Comptonization models mentioned above.

The first significant difference is the fact that our model is intrinsically
 non-stationary and the evolution in time of the resulting spectra (as it has been  
shown in Section~4.2) turns out
to be a decisive factor in order to reproduce the observed  steepening of 
the X-ray spectrum above 100~keV, in accordance with OSSE and BeppoSAX results.

Secondly,  the compactness parameter (see below for its explicit definition, as given by 
Guilbert
et al. \cite{guilbert}), consistently evaluated for the coronal emitting regions 
in our scenario, is lower than the critical limit for pair production to be significant, thus 
avoiding, in our case, the problem of the annihilation line, generally expected from standard, stationary 
non-thermal models, but not detected by OSSE (Johnson et al. \cite{johnson97}; Zdziarski et al.
\cite{zdz97}).  
In fact, following Guilbert et al. (\cite{guilbert}),
the compactness parameter  of the coronal structure can be estimated as
 $l_{\rm corona} \sim \sigma_{\rm T}/(m_{\rm e}c^3) (L_{\rm X}/R_{\rm UV-X})$, where $L_{\rm X}$
 is the source X-ray luminosity and $R_{\rm UV-X}$  is
the global extent of the region where the IC reprocessing occurs.
 Using  for $L_{\rm X}$ values in the range $2-3\times 10^{43}$~erg/s (as reported by
 Nicastro et al. \cite{nica}, Perola et al. \cite{perola02}, Bianchi et al.
 \cite{bianchi04} for NGC~5548), we get an estimate
 $l_{\rm corona} \sim 4.9-7.4$, reasonably below the well known limit for pair
 production to become effective, namely $l\ga 4\pi$.
Even lower values can be estimated
for the compactness parameter for the single emitting loop 
(the ``direct'' analogue of the ``blob'' of Haardt
et al. \cite{hmg94}), since in that case
 we have $l_{\rm loop}\sim \sigma_{\rm T}/(m_{\rm e}c^3)(L_{\rm X}/ q)/R_{\rm em}\sim
 2.3-3.4$, to be compared  with the reference value $\sim 30$ of
 Haardt et al. (\cite{hmg94}).  Therefore, in our framework
neglecting pair production/annihilation effects on the spectrum turns out to be  fully
justified from the resulting conditions of the emitting structure, and 
 this supports the consistency of our treatment. 
It is also noteworthy 
that this condition differentiates our model from the thermal Comptonization ones as well, as it 
is clear from the comparison above with compactness properties of
 the thermal ``patchy-corona'' model of Haardt et al. (\cite{hmg94}).

Still referring to the issue of a comparison of our presently proposed scenario with preexistent 
models, we have to mention that models including both a thermalized (Compton efficient) electron 
population and a non-thermal relativistic component have been also devised; these models have 
been developed to account for the fact that a purely thermal (Maxwellian) distribution for 
energetic electrons is quite difficult to achieve (see Coppi \cite{coppi99}). 

On one hand thermalization
mechanisms, which are more efficient than Coulomb collision for energy exchange, 
have been explored, although, according to Coppi (\cite{coppi99}), 
``realistic calculations are in general quite difficult, especially near an accreting black hole,
where we still have relatively poor knowledge of the exact physical conditions''.
It is far beyond the scope of the present paper to analyze this topic in detail; we just 
mention one interesting model belonging to this first sub-class, namely the so-called ``synchrotron
boiler'', devised by Ghisellini et al. (\cite{ggs88}) and further developed by Ghisellini
et al. (\cite{ghs98}), based on efficient self-absorption of
synchrotron photons, emitted by electrons in the source, by different electrons of the same
population, thus ``rapidly'' exchanging energy and producing relaxation of the particle 
population itself to a resulting steady-state quasi-Maxwellian distribution, with a high-energy tail. 
It is important to notice that the main assumptions for this mechanism to  be efficient, and thus
for thermalization to effectively occur (Ghisellini et al. \cite{ghs98}),
are that i) the mean energy, $<\gamma>$, of the energetic electrons is just a few, and 
ii) the magnetic energy density is dominant with respect to the radiation energy density,
together with the further condition that the soft photons to be Comptonized
{\it only} arise from the reprocessing of about half of the hard radiation by cold (disk) matter
close to the active (Comptonizing) region. 
None of these conditions is met in our scenario, in which no such straightforward and stringent 
relation between the external soft radiation energy density and the hard (i.e. X-ray) one is  
required,  thus allowing our magnetic energy density to be somewhat lower
than the {\it soft} seed radiation energy density, and in any case not dominant 
(even though magnetic energy is in our view the reservoir 
for the electron impulsive energization, and, therefore, ultimately, for the high energy emission);
moreover, contrary to assumption i), mentioned above, for the thermalization mechanism to 
be efficient, 
the accelerated electrons we consider are characterized by much higher electron energy
($\gamma_{\rm min} \gg 1$).

On the other hand,  models intrinsically allowing for  a ``lower'' energy portion of the electron 
distribution of a Maxwellian form and a steady injection of high energy electrons in a
power-law or delta-function distribution  have also been devised (Coppi \cite{coppi99}). These are
the so-called hybrid thermal/non-thermal models and the  resulting steady state distribution 
is in fact a Maxwellian plus a high energy non-thermal tail, whose relative importance depends
basically on the ratio of the power supplied to the thermalized component (heating rate) and the
power injected in non-thermal electrons ($l_{\rm th}/l_{\rm nth}$ in terms of compactnesses)
as well as on the non-thermal injection spectrum
(especially the characteristic $\gamma$ range). As a consequence, the resulting high energy
spectrum depends on these parameters  as well. It is well known (see Ghisellini
et al. \cite{ghf93}) that, as long as the radiative process is multiple Compton scattering
and the maximum particle energy is a few MeV, the resulting high energy photon spectrum is
indistinguishable from the thermal expected one, and is basically independent of the details of the
distribution of particles. When the $\gamma_{\rm max}$ of the non-thermal electrons is instead 
quite high, the spectral properties will depend on the ratio $l_{\rm th}/l_{\rm nth}$, but, even 
when this ratio is $\gg 1$, the spectrum should be distinguishable from the thermal
one at gamma-ray energies. These models are at present mostly applied to  galactic black holes
in their soft states,
that cannot be fitted by the thermal models,  since the corresponding spectra do show power-law
like tails at least up to 1~MeV (although relatively weak), and, as a consequence, appear to
require a high energy non-thermal tail in the electron distribution contributing to the formation 
of the spectrum itself; 
on the contrary, for X-ray spectra of normal broad-line Seyfert~1 active nuclei, 
thermal Comptonization models have been preferred, up to now.
It is appropriate, here, to cite a significant statement in Coppi's (\cite {coppi99}) review on 
hybrid models, again enlightening a substantial difference with our presently proposed scenario:  
the spectra produced by Coppi's code for hybrid models are 
{\it ``steady state, while the real spectra that are being fit are typically time integrations over 
many flares...''}. 

In the light of all the arguments above discussed, we can conclude that 
while the comparison of our model with observations
 suggests that our model
 is plausible and can be considered a valid alternative to
thermal  Comptonization models, a deeper analysis
is still necessary to fully understand the physics of the emitting medium in AGN coronae.

\begin{acknowledgements}
This work was partially supported by the Italian Ministry of Research
(MIUR).
\end{acknowledgements}


\begin{thebibliography}{}


\bibitem[2004]{bianchi04}
Bianchi, S., Matt, G., Balestra, I., Guainazzi, M.,\& Perola, G.C. 2004, A\&A 422, 65

\bibitem[1970]{blum}
Blumenthal, G.R., \& Gould, R.J. 1970, Rev. Mod. Phys., 42, 237

\bibitem[2003]{blustin}
Blustin, A.J., Branduardi-Raymont, G., Behar, E., et al. 2003, A\&A 403, 481

\bibitem[1997a]{boella97a}
Boella G., Butler R., Perola G., et al. 1997a, A\&AS, 122, 299

\bibitem[1997b]{boella97b}
Boella G., Chiappetti L., Conti G., et~al. 1997b, A\&AS, 122, 327

\bibitem[2002]{broth}
Brotherton, M.S., Green, R. F., Kriss, G. A., et al. 2002, ApJ, 565, 800

\bibitem[1987]{carle}
Carleton, N.P., Elvis, M., Fabbiano, G., et al. 1987, ApJ, 318, 595
 
\bibitem[2000]{chiang}
Chiang, J., Reynolds, C. S., Blaes, O. M., et al. 2000, ApJ, 528, 292

\bibitem[1999]{coppi99}
Coppi, P. S. 1999,  in High Energy Processes In Accreting Black Holes, ed. J. Poutanen \& 
R. Svensson (San Francisco: ASP), ASP Conf. Ser. 161, 375

\bibitem[1998]{dima98}
Di Matteo, T. 1998, MNRAS, 299, L15

\bibitem[1997]{dove}
Dove, J.B., Wilms, J., \& Begelman, M.C. 1997, ApJ, 487, 747

\bibitem[1999]{fiore}
Fiore F., Guainazzi M., \& Grandi P. 1999, Cookbook for {\t BeppoSAX}  Spectral Analysis

\bibitem[1997]{frontera}
Frontera F., Costa E., Dal Fiume D., et~al. 1997, A\&AS, 122, 371

\bibitem[1979]{ga79}
Galeev, A.A., Rosner, R., \& Vaiana, G.S. 1979, ApJ, 229, 318

\bibitem[1988]{ggs88}
Ghisellini, G., Guilbert, P., \& Svensson, R. 1988, ApJL, 335, L5 

\bibitem[1993]{ghf93}
Ghisellini, G., Haardt, F., \& Fabian, A.C. 1993, MNRAS, 263, L9 
 
\bibitem[1994]{ghise94}
Ghisellini, G. 1994, in ``Frontiers of Space and Ground-Based Astronomy'', eds. Wamsteker, W., 
Longair, M.S., \& Kondo, Y., p.~347, Kluwer Academic Publishers, Dordrecht: The Netherlands 
 
\bibitem[1998]{ghs98}
Ghisellini, G., Haardt, F., \& Svensson, R. 1998, MNRAS, 297, 348
 
\bibitem[2004]{ghise04}
Ghisellini, G., Haardt, F., \& Matt, G. 2004, A\&A, 413, 535

\bibitem[1969]{gin}
Ginzburg , V.L. 1969, Elementary processes for cosmic ray astrophysics,
  Gordon and Breach  

\bibitem[1996]{gondek96}
Gondek, D., Zdziarski, A. A., Johnson, W. N., et al. 1996, MNRAS, 282, 646

\bibitem[1983]{guilbert}
Guilbert P.W., Fabian A.C., \& Rees M.J. 1983, MNRAS, 205,593

\bibitem[1991]{hm91}
Haardt, F., \& Maraschi L. 1991, ApJL, 380, L51

\bibitem[1993]{hm93}
Haardt, F., \& Maraschi L. 1993, ApJ, 413, 507

\bibitem[1994]{hmg94}
 Haardt, F.,  Maraschi, L. \& Ghisellini, G. 1994, ApJL, 432, L95

\bibitem[1997]{haardt97}
Haardt, F. 1997, Mem. Soc. Astron. It., 68, 73

\bibitem[2003]{haba}
Haba, Y., Kunieda, H., Misaki, K., et al. 2003, ApJ, 599, 949

\bibitem[1989]{hp89}
Heyvaerts, J.F., \& Priest, E.R. 1989, A\&A, 216, 230

\bibitem[1997]{johnson97}
Johnson, W. N., McNaron-Brown, K., Kurfess, J. D., et al. 1997, ApJ, 482, 173 

\bibitem[2000]{kaastra00}
Kaastra, J.S., Mewe, R., Liedhal, D. A., Komossa, S., \& Brinkmann, A. C. 2000, A\&A, 354, L83

\bibitem[2002]{kaastra}
Kaastra, J.S., Steenbrugge, K. C., Raassen, A. J. J., et al. 2002, A\&A, 386, 427

\bibitem[2004]{kaas2004}
Kaastra, J.S., Steenbrugge, K. C., Crenshaw, D. M., et al. 2004, A\&A 422, 97

\bibitem[1997]{kl97}
Klein, K.-L., Aurass, H., Soru-Escaut, I.,  \& Kalman, B. 1997, A\&A, 320, 612

\bibitem[1999]{korat99}
Koratkar, A., \& Blaes, O. 1999, PASP, 111, 1

\bibitem[1999]{krolik99}
Krolik, J.H. 1999, Active Galactic Nuclei, Princeton University Press

\bibitem[1980]{lang}
Lang, K.R. 1980, Astrophysical Formulae, Springer Verlag 

\bibitem[1997]{lesch}
Lesch, H., \& Birk, G.T. 1997, A\&A, 324, 461

\bibitem[1977]{liang77} 
Liang, E.P.T, \& Price, R.H. 1977, ApJ, 284, 247

\bibitem[1979]{liang79} 
Liang, E.P.T. 1979, ApJL, 231, L111

\bibitem[2002]{liu02}
Liu, B.F.,Mineshige, S., \& Shibata, K. 2002, ApJL, 572, L173


\bibitem[2004]{hardy}
McHardy, I.M., Papadakis, I. E., Uttley, P., Page, M. J., \&  Mason, K. O.  2004, MNRAS, 348,783

\bibitem[1995]{magdz}
Magdziarz, P., \&  Zdziarski, A.A. 1995, MNRAS, 273, 837

\bibitem[2001]{mark01}
Markowitz, A., \& Edelson, R. 2001, ApJ 547, 684

\bibitem[2003]{mark03}
Markowitz, A.,  Edelson, R., \& Vaughan, S. 2003, ApJ 598, 935

\bibitem[1982]{malkan}
Malkan, A.M., \&  Sargent, W.L.W. 1982, ApJ, 254, 23

\bibitem[1994]{masu94}
Masuda, S.,  Kosugi, T., Hara, H., Tsuneta, S., \& Ogawara, Y. 1994, Nature, 371, 495

\bibitem[2001]{matt01}
Matt, G. 2001, in AIP Conf. Proc. 599, X-Ray Astronomy: Stellar Endpoints, AGN, and the Diffuse
X-Ray Background, eds. N. E. White, G. Magaluti, \& G. Palumbo (New York: AIP), 209 

\bibitem[2001]{merlo}
Merloni, A., \&  Fabian, A.C. 2001, MNRAS, 321, 549

\bibitem[1976]{melro76}
Melrose, D.B., \& Brown, F.C. 1976, MNRAS, 176, 15

\bibitem[2000]{misto00}
Miller, K.A., \& Stone, J. M. 2000, ApJ, 534, 398

\bibitem[1994]{nandra94}
Nandra, K., \& Pounds, K.A. 1994, MNRAS, 268, 405

\bibitem[1997b]{nandra97bis}
Nandra, K., George, I. M., Mushotzky, R. F., Turner, T. J., \& Yaqoob, T.  1997b, ApJ, 476, 70

\bibitem[1997a]{nandra97}
Nandra, K.,  George, I. M., Mushotzky, R. F., Turner, T. J., \& Yaqoob, T. 1997a, ApJ, 477, 602

\bibitem[1999a]{nica2}
Nicastro, F.,  Fiore, F., Perola, G. C., Elvis, M. 1999a, ApJ, 512, 184

\bibitem[1999b]{nica3}
Nicastro, F., Fiore, F., \& Matt, G. 1999b, ApJ, 517, 108

\bibitem[2000]{nica}
Nicastro, F., Piro, L., De Rosa, A., et al. 2000, ApJ, 536, 718

\bibitem[1997]{parmar}
Parmar A.N., Martin D.D.E., Bavdaz M., et~al. 1997, A\&AS, 122, 309

\bibitem[2002]{perola02}
Perola, G.C., Matt, G., Cappi, M., et al. 2002, A\&A, 389, 802

\bibitem[2000]{petru}
Petrucci, P.O., Haardt, F., Maraschi, L., et al. 2000, ApJ, 540, 131

\bibitem[2003]{pounds}
Pounds, K., Reeves, J. N., Page, K. L., et al. 2003, MNRAS, 341, 953 

\bibitem[1996]{pout}
Poutanen, J., \&  Svensson, R. 1996, ApJ, 470, 249

\bibitem[2004]{risaliti04}
Risaliti, G., \& Elvis, M. 2004, astro-ph/0407291

\bibitem[1998]{schop}
Schopper, R., Lesch, H., \&  Birk, G.T. 1998, A\&A, 335, 26 
 
\bibitem[1973]{ss73}
Shakura, N.I., \& Sunyaev, R.A. 1973, A\&A, 24, 337
 
\bibitem[1989]{shibata89}
Shibata, K., Tajima, T., Steinolfson, R.S., \& Matsumoto, R. 1989, ApJ, 345, 584

\bibitem[2003]{steen}
Steenbrugge, K.C., Kaastra, J. S., de Vries, C. P., \&  Edelson, R. 2003, A\&A, 402, 477

\bibitem[1994]{sv94}
Svensson, R. 1994, ApJS, 92, 585 

\bibitem[1996]{sv96}
Svensson, R. 1996, A\&AS, 120, 475 

\bibitem[2003]{uttley}
Uttley, P., Edelson, R., McHardy, I. M., Peterson, B. M., \& Markowitz, A. 2003, ApJL, 584, L53

\bibitem[1990]{wamst}
Wamsteker, W., Rodriguez-Pascual, P., Wills, B. J., et al. 1990, ApJ, 354, 446

\bibitem[2000]{wande}
Wandel, A., Peterson, B.M., \&  Malkan, M.A. 2000, ApJ, 526, 579

\bibitem[1987]{ward}
Ward, W., Elvis, M., Fabbiano, G., et al. 1987, ApJ, 315, 74

\bibitem[1995]{zdz95}
Zdziarski, A.A., Johnson, W. N., Done, C., Smith, D., \&  McNaron-Brown, K. 1995, ApJL, 438, L63

\bibitem[1997]{zdz97}
Zdziarski, A.A., Johnson, W.N., Poutanen, J., Magdziarz, P., \& Gierlinski, M. 1997, in 
The Transparent Universe, Proceedings of the 2nd INTEGRAL Workshop, eds. C. Winkler, T.J.-L. 
Courvoisier, \& P. Durouchoux, ESA-SP 382, Noordwijk: ESA, 373

\bibitem[2000]{zdz00}
Zdziarski, A.A., Poutanen, J., \& Johnson, W.N. 2000, ApJ, 542, 703



\end{thebibliography}
\end{document}